\documentclass[a4paper,dvips,11pt]{article}

\usepackage{cite}
\usepackage{epsf}
\usepackage{multirow} 
\usepackage{bm,bbm} 
\usepackage[pdftex]{graphicx}
\usepackage{tabularx}
\usepackage{latexsym}
\usepackage{amsmath,amssymb,exscale}
\usepackage{array,multicol}
\numberwithin{equation}{section}

\def\id{\ 1 \! \! \! \! 1}
\def\draftlabel#1{{\@bsphack\if@filesw {\let\thepage\relax
   \xdef\@gtempa{\write\@auxout{\string
      \newlabel{#1}{{\@currentlabel}{\thepage}}}}}\@gtempa
   \if@nobreak \ifvmode\nobreak\fi\fi\fi\@esphack}
        \gdef\@eqnlabel{#1}}
\def\@eqnlabel{}
\def\@vacuum{}
\def\draftmarginnote#1{\marginpar{\raggedright\scriptsize\tt#1}}
\def\draft{\oddsidemargin -.5truein
        \def\@oddfoot{\sl preliminary draft \hfil
        \rm\thepage\hfil\sl\today\quad\militarytime}
        \let\@evenfoot\@oddfoot \overfullrule 3pt
        \let\label=\draftlabel
        \let\marginnote=\draftmarginnote
   \def\@eqnnum{(\theequation)\rlap{\kern\marginparsep\tt\@eqnlabel}%
\global\let\@eqnlabel\@vacuum}  }

\newcommand{\PRL}[3]{\emph{ Phys.~Rev.~Lett.} \textbf{#1} (#2) #3}

\newcommand{\PR}[3]{\emph{ Phys.~Rep.} \textbf{#1} (#2) #3}

\def\ov{\overline}

\def\dalemb#1#2{{\vbox{\hrule height .#2pt
         \hbox{\vrule width.#2pt height#1pt \kern#1pt
                 \vrule width.#2pt}
         \hrule height.#2pt}}}

\def\half{{\textstyle{1\over2}}}
\let\a=\alpha    
    
    \let\p=\pi 
\let\s=\sigma     

      \let\G=\Gamma  
  \let\S=\Sigma   
\let\F=\Phi

 \def\bd{\begin{document}} \def\ed{\end{document}}
\def\ds{\documentstyle} \let\fr=\frac \let\bl=\bigl \let\br=\bigr
\let\Br=\Bigr \let\Bl=\Bigl
\let\bm=\bibitem
\let\na=\nabla
\let\pa=\partial
\let\ov=\overline
\def\ie{{\it i.e.\ }}
\def\tr{{\mbox{\rm tr}}}
\newcommand{\be}{\begin{equation}}
\newcommand{\ee}{\end{equation}}
\newcommand{\beba}{\begin{equation}\begin{array}{lcl}}
\newcommand{\eaee}{\end{array}\end{equation}}
\newcommand{\bea}{\begin{eqnarray}}
\newcommand{\eea}{\end{eqnarray}}
\newcommand{\ba}{\begin{array}}
\newcommand{\ea}{\end{array}}
\newcommand{\td}{\tilde}
\newcommand{\norsl}{\normalsize\sl}
\newcommand{\ns}{\normalsize}
\newcommand{\refs}[1]{(\ref{#1})}
\def\simlt{\mathrel{\lower2.5pt\vbox{\lineskip=0pt\baselineskip=0pt
            \hbox{$<$}\hbox{$\sim$}}}}
\def\simgt{\mathrel{\lower2.5pt\vbox{\lineskip=0pt\baselineskip=0pt
            \hbox{$>$}\hbox{$\sim$}}}}
\def\A{{\cal A}}
\def\a{{\mathcal a}}
\def\V{{\cal V}}
\def\F{{\cal F}}
\def\p{{\mathcal \phi}}
\def\L{{\mathcal L}}
\def\M{{\mathcal M}}
\def\bD{{\ov {\rm D}}}
\def\bO{{\ov {\rm O}}}
\def\bOp{{\ov {\rm O'}}}
\def\O{{ {\rm O}}}

\newcommand{\nsect}{\setcounter{equation}{0}
\def\theequation{\thesection.\arabic{equation}}\section}
\newcommand{\nappend}{\setcounter{equation}{0}
\def\theequation{\rm{A}.\arabic{equation}}\section*}
\newcommand{\appendixA}{\setcounter{equation}{0}
\def\theequation{\rm{A}.\arabic{equation}}\section*}
\newcommand{\appendixB}{\setcounter{equation}{0}
\def\theequation{\rm{B}.\arabic{equation}}\section*}
\newcommand{\appendixC}{\setcounter{equation}{0}
\def\theequation{\rm{C}.\arabic{equation}}\section*}
\newcommand{\appendixD}{\setcounter{equation}{0}
\def\theequation{\rm{D}.\arabic{equation}}\section*}
\newcommand{\appendixE}{\setcounter{equation}{0}
\def\theequation{\rm{E}.\arabic{equation}}\section*}
\newcommand{\appendixF}{\setcounter{equation}{0}
\def\theequation{\rm{F}.\arabic{equation}}\section*}
\newcommand{\appendixG}{\setcounter{equation}{0}
\def\theequation{\rm{G}.\arabic{equation}}\section*}
\def\baselinestretch{1.5}
\catcode`\@=11
\def\marginnote#1{}
\newcount\hour
\newcount\minute
\newtoks\amorpm
\hour=\time\divide\hour by60
\minute=\time{\multiply\hour by60 \global\advance\minute by-\hour}
\edef\standardtime{{\ifnum\hour<12 \global\amorpm={am}%
        \else\global\amorpm={pm}\advance\hour by-12 \fi
        \ifnum\hour=0 \hour=12 \fi
        \number\hour:\ifnum\minute<10 0\fi\number\minute\the\amorpm}}
\edef\militarytime{\number\hour:\ifnum\minute<10 0\fi\number\minute}
\def\draftlabel#1{{\@bsphack\if@filesw {\let\thepage\relax
   \xdef\@gtempa{\write\@auxout{\string
      \newlabel{#1}{{\@currentlabel}{\thepage}}}}}\@gtempa
   \if@nobreak \ifvmode\nobreak\fi\fi\fi\@esphack}
        \gdef\@eqnlabel{#1}}
\def\@eqnlabel{}
\def\@vacuum{}
\def\draftmarginnote#1{\marginpar{\raggedright\scriptsize\tt#1}}
\def\draft{\oddsidemargin -.5truein
        \def\@oddfoot{\sl preliminary draft \hfil
        \rm\thepage\hfil\sl\today\quad\militarytime}
        \let\@evenfoot\@oddfoot \overfullrule 3pt
        \let\label=\draftlabel
        \let\marginnote=\draftmarginnote
   \def\@eqnnum{(\theequation)\rlap{\kern\marginparsep\tt\@eqnlabel}%
\global\let\@eqnlabel\@vacuum}  }

\def\preprint{\twocolumn\sloppy\flushbottom\parindent 1em
        \leftmargini 2em\leftmarginv .5em\leftmarginvi .5em
        \oddsidemargin -.5in    \evensidemargin -.5in
        \columnsep 15mm \footheight 0pt
        \textwidth 250mmin      \topmargin  -.4in
        \headheight 12pt \topskip .4in
        \textheight 175mm
        \footskip 0pt
        \def\@oddhead{\thepage\hfil\addtocounter{page}{1}\thepage}
        \let\@evenhead\@oddhead \def\@oddfoot{} \def\@evenfoot{} }

\def\titlepage{\@restonecolfalse\if@twocolumn\@restonecoltrue\onecolumn
     \else \newpage \fi \thispagestyle{empty}\c@page\z@ 
        \def\thefootnote{\fnsymbol{footnote}} }

\def\endtitlepage{\if@restonecol\twocolumn \else  \fi
        \def\thefootnote{\arabic{footnote}}
        \setcounter{footnote}{0}}  

\catcode`@=12
\relax
\def\abs#1{\left| #1\right|}
\def\bC{\mathop{\bf C}}
\def\bea{\begin{array}}
\def\bem{\begin{displaymath}}
\def\beq{\begin{equation}}
\def\bea{\begin{eqnarray}}
\def\bR{\mathop{\bf R}}
\def\bra#1{\left\langle #1\right|}
\def\eea{\end{array}}
\def\eem{\end{displaymath}}
\def\eeq{\end{equation}}
\def\eea{\end{eqnarray}}
\def\eq{\beq\eeq}                          
\def\eqr#1{\beq\label#1\eeq}               
\def\half{\frac{1}{2}}
\def\Im{\mathop{\rm Im}}
\def\ket#1{\left| #1\right\rangle}
\def\sket#1{| #1 >}
\def\lie{\hbox{\it \$}}                          
\def\lineint{\oint \frac{d z}{2 \pi i}} 
\def\modsq#1{| #1 |^2}
\def\NP#1#2#3{Nucl. Phys. \underline{#1} (19#2) #3}
\def\ov{\overline}
\def\partder#1#2{{\partial #1\over\partial #2}}
\def\PL#1#2#3{Phys. Lett. \underline{#1} (19#2) #3}
\def\PR#1#2#3{Phys. Rev. \underline{#1} (19#2) #3}
\def\PRL#1#2#3{Phys. Rev. Lett. \underline{#1} (19#2) #3}
\def\Re{\mathop{\rm Re}}
\def\secder#1#2#3{{\partial^2 #1\over\partial #2 \partial #3}}
\def\s2w{\sin^2 \theta_W}
\def\Tr{\mathop{\rm Tr}}
\def\und{\underline}
\def\VEV#1{\left\langle #1\right\rangle} \let\vev\VEV
\def\mbf#1{\hbox{\boldmath $#1$}}
\relax
\def\dalpha{{\dot\alpha}}
\def\dbeta{{\dot\beta}}
\def\drho{{\dot\rho}}
\def\dsigma{{\dot\sigma}}
\def\crbig{\\\noalign{\vspace {3mm}}}
\def\bigint{{\displaystyle\int}}
\def\S{\Sigma}
\def\G{\Gamma}
\def\L{{\cal L}}
\def\SG{S_{\Gamma}}
\def\Fint{{\bigint d^2\theta\,}}
\def\Fbarint{{\bigint d^2\ov\theta\,}}
\def\Dint{{\bigint d^2\theta d^2\ov\theta\,}}

\title{ \vspace*{-0.8cm}
\begin{flushright}
\normalsize{CERN-PH-TH/2006-216\\ 
CPTH-RR-079-1006\\
IP/BBSR/2006-18\\
\texttt{hep-th/0610246}}\\
\end{flushright}
\vspace{1cm}
\bf{Magnetic fluxes and moduli stabilization} \vspace*{-0.3cm}}

\date{}
\begin{document}

\author{\bf\large{
Ignatios Antoniadis$^{1,2}$
~\footnote{Ignatios.Antoniadis@cern.ch}~,
Alok Kumar$^{3}$\footnote{kumar@iopb.res.in}~, 
Tristan Maillard$^{1,2,4}$\footnote{Tristan.Maillard@cpht.polytechnique.fr}}\\  
\\[-3mm]
\emph{\normalsize $^1$Department of Physics, CERN - Theory Division, 
CH-1211 Geneva 23, Switzerland }\\
\emph{\normalsize $^2$ CPHT, UMR du CNRS 7644 , Ecole Polytechnique, 91128 Palaiseau, France}\\
\emph{\normalsize $^3$Institute of Physics, Bhubaneswar 751 005, 
India}\\
\emph{\normalsize $^4$Institut f\"ur Theoretische Physik, ETH 
H\"onggerberg, CH-8093\, Z\"urich, Switzerland}
}

\date{}

\maketitle
\thispagestyle{empty}

\begin{abstract}
Stabilization of closed string moduli in toroidal orientifold
compactifications  of type IIB string theory
are studied using constant internal magnetic fields 
on $D$-branes  and 3-form fluxes 
that preserve ${\cal N}=1$ supersymmetry in four dimensions.
Our analysis corrects and extends previous work by us, and
indicates that charged scalar VEV's need to be turned on, 
in addition to the fluxes, in order to construct a consistent supersymmetric model.
As an explicit example, we first show the stabilization of all K\"ahler 
class and complex structure moduli 
by turning on magnetic fluxes on different sets of $D9$-branes 
that wrap the internal space $T^6$ in a compactified type I string theory,
when a charged scalar on one of these branes acquires a non-zero VEV. 
The latter can also be determined by adding extra magnetized branes,
as we demonstrate in a subsequent example.
In a different model with magnetized $D7$-branes, 
in a IIB orientifold on $T^6/\mathbb{Z}_2$, we show the
stabilization of all the closed string moduli, including the axion-dilaton
at weak string coupling $g_s$, by turning on appropriate closed
string 3-form fluxes.
\end{abstract}

\date


\maketitle
 \vspace*{-0.8cm}
\hrulefill
\newpage
\section{Introduction}
String theory is known to possess a large number of vacua which 
contain the basic structure of grand unified theories, and in 
particular of the Standard Model. 
However, one of the major stumbling blocks in making 
further progress along these lines has been the lack of a guiding
principle for choosing the true ground state of the theory, thus 
implying the loss of predictivity.
In particular, string vacua depend in general on continuous 
parameters, characterizing for instance the size and shape of the
compactification manifold,
that correspond to vacuum expectation values (VEV's) of the 
so-called moduli fields. These are perturbative flat directions of 
the scalar potential, at least as long as supersymmetry remains unbroken.
It is therefore of great interest that during the last few years
there has been a considerable success in fixing the string ground 
states, by invoking principles similar to the spontaneous 
symmetry breaking mechanism, now in the context of string theory.
In particular, it has been realized that closed, as well as open,
string background fluxes can be turned on,
fixing the VEV's of the moduli fields and therefore
providing the possibility for choosing a ground state as a local
isolated minimum of the scalar potential of the theory.
This line of approach allows string theory to play directly a role 
in particle unification, predicting the strength of interactions and
the mass spectrum. In particular, the string coupling becomes a
calculable dynamical parameter that fixes the value of the fine
structure constant and determines the Newtonian coupling in 
terms of the string length.

On one hand, moduli stabilization using closed string 
3-form fluxes has been discussed in a great detail in the 
literature \cite{Frey:2002hf, Review}. 
${\cal N}=1$ space-time supersymmetry and
various consistency requirements imply that the 
3-form fluxes must satisfy the following conditions formulated
on the complexified flux defined as $G = F - \phi H $, 
where $F$ and $H$ are the R-R (Ramond) and
NS-NS (Neveu-Schwarz) 3-forms, respectively, 
and $\phi$ is the axion-dilaton modulus: 
(1) The only non-vanishing components of $G$ are of the type $(2,1)$, 
pointing along two holomorphic and one anti-holomorphic directions,
implying that its $(1,2)$, $(3,0)$ and $(0,3)$ components
are zero; (2) $G$ is primitive, requiring $J \wedge G = 0$
with $J$ being the K\"ahler form. 
This approach has been applied to orientifolds of both 
toroidal models as well as of Calabi-Yau compactifications. 
However, a drawback of the method is that the K\"ahler class 
moduli remain undetermined due to the absence of a harmonic $(1,0)$ 
form on Calabi-Yau spaces, implying that the constraint 
$J \wedge G = 0$ is trivially satisfied. 
In the toroidal orientifold case, it turns out that one is
able to stabilize the K\"ahler class moduli only partially,
but in particular the overall volume remains always unfixed.

On the other hand, in \cite{AM} two of the present authors have 
shown that both complex structure and K\"ahler class moduli may 
be stabilized in the type I string theory compactified down to four 
dimensions.\footnote{For partial K\"ahler moduli stabilization, see 
also \cite{Blumenhagen:2003vr, Cascales:2003zp}.} 
This can be achieved by turning on magnetic fluxes which 
couple to various $D9$-branes, that wrap on $T^6$, through a 
boundary term in the open string world-sheet action. The latter 
modifies 
the open string Hamiltonian and its spectrum, and puts 
constraints on the closed string background fields due to their 
couplings to the open string action. 
More precisely, supersymmetry conditions in the presence of 
branes with magnetic fluxes, together with conditions which define 
a meaningful (string) theory, put restrictions on the values of 
the moduli and fix them to specific constant values.
This also  breaks the original ${\cal N}=4$ supersymmetry of the 
compactified type I theory to an ${\cal N}=1$ supersymmetric gauge 
theory with a number of chiral multiplets. A detailed analysis of the 
final spectrum, as well as other related issues have been discussed 
in \cite{Bianchi:2005yz}. 

In the simplest case, the above model has only $O9$ orientifold 
planes and several stacks of magnetized $D9$-branes. The main
ingredients for moduli stabilization are then: (1) the introduction
of ``oblique" magnetic fields, needed to fix the off-diagonal
components of the metric, that correspond to mutually 
non-commuting matrices similar to non-abelian orbifolds;
(2) the property that magnetized $D9$-branes can lead to negative
5-brane tensions; and (3) the non-linear part of Dirac-Born-Infeld 
(DBI) action which is needed to fix the overall volume.
Actually, the first two ingredients are also necessary for
satisfying the 5-brane tadpole cancellation without adding
$D5$-branes or $O5$ planes, while the last two properties
are only valid in four-dimensional compactifications (and not
in higher dimensions). It turns out however that the conditions
of supersymmetry and tadpole cancellation cannot be satisfied
simultaneously in such simple setups, contrary to our previous 
claim.\footnote{This is due to an overlooked sign in \cite{AM},
as explained in Section 2.}

In this work, we point out that an additional ingredient,
namely  non-vanishing VEV's for the scalars fields on
some of the branes, is needed for consistent model building. 
Indeed, it is known that the brane fluxes generate 
a $D$-term potential for the K\"ahler moduli in the form of a
Fayet-Iliopoulos (FI)-term. The supersymmetry
conditions used in \cite{AM} are then precisely the $D$-flatness 
conditions, implying the vanishing of the FI-parameters. But this is valid 
only in the case where the VEV's of the charged scalars vanish. In their 
presence, the $D$-flatness condition for the closed string moduli 
is modified and a more general supersymmetric minimum can be
obtained. 

In this paper, we first implement this new ingredient on the models of 
\cite{AM}. We show that by a minimal modification of those models, namely
a modification in the supersymmetry condition on one of the branes
by switching a non-zero VEV $v$, the results of \cite{AM}
still hold. In other words, one is able  to stabilize the complex structure 
and K\"ahler class moduli, while cancelling 5-brane tadpoles among the 
magnetized branes, and $D9$-brane 
tadpoles by the $O9$ planes of type I string theory. Furthermore,
this method of moduli stabilization can be extended for the 
$T^6/\mathbb{Z}_2$ orientifold compactification of IIB string theory.
One can then try to stabilize also the dilaton-axion modulus
by turning on simultaneously 3-form fluxes. Since in their presence,
the gauge groups of the $D9$-branes are in general anomalous~\cite{Freed:1999vc}, 
we construct models with magnetized $D7$-branes.
In this case, we explicitly show that all closed string moduli are now
fixed using brane fluxes, 3-form closed string fluxes and charged scalar
VEV's on some of the branes.
 
Stabilization of all closed string moduli, using magnetized 
$D9$-branes, were attempted previously by us in \cite{AKM}.
However due to the use of an inappropriate definition of wrapping
numbers for oblique fluxes, erroneous tadpole contributions
were obtained. In this paper, we have corrected the error
and obtained a model with tadpole cancellations, by
turning on charged scalar VEV's, where all the closed string
moduli are fixed.

Our moduli stabilization scheme, where some of the open string scalars
have acquired VEV's to generate a non-zero FI parameter, can 
be thought of as a mixing of 
K\"ahler moduli with open string fields through the D-term, in such a way that only
one linear combination is fixed by the presence of the corresponding 
magnetic field \cite{Antoniadis:1997mm}, while its orthogonal combination
remains a flat direction.
However, we show that the latter can also be fixed by the same principle,
implying a particularly interesting algorithm for moduli stabilization in
toroidal type I compactifications: (1) All geometric moduli are first fixed
using a minimal set of (at least nine) magnetized branes, in the absence
of charged scalar VEV's, in the spirit of \cite{AM}. This has the advantage
of being exact in $\alpha'$ (world-sheet) perturbation theory, but does not
satisfy tadpole cancellation. (2) The latter is achieved by adding extra
magnetized branes on which some charged scalars are forced to acquire 
non-vanishing VEV's. Since the inclusion of charged fields in the D-terms
is not known exactly, their VEV's can be determined only perturbatively
in $\alpha'$, when their values are small compared to the string scale.

The supersymmetry conditions, as well as the tadpole equations, possess
some scaling symmetries observed in \cite{AM}, where fluxes and K\"ahler
moduli are rescaled appropriately for fixed charged scalar VEV's. These 
symmetries lead to an infinite discrete class of vacua with stabilized moduli,
differing by their spectra and values of gauge couplings and Planck mass
in string units.

The rest of the paper is organized as follows. In Section 2, we 
write down the consistency conditions for magnetic fluxes on 
$D9$-branes in the context of type I toroidal 
compactifications. Explicit models are then presented in Section 3.
The first model is a minimal modification of the main example
of \cite{AM}, illustrating our method of moduli stabilization in
the presence of a single VEV for one of the brane stacks.
It also serves as an example to demonstrate the existence of the
discretum of vacua and to discuss models with large dimensions.
In a second model, we show how the open string scalar VEV's
are fixed consistently by adding extra stacks of magnetized branes.
Consistency conditions, as well as a model with stabilized moduli,
for magnetized $D7$-branes in a IIB compactification on $T^6/\mathbb{Z}_2$
is given in Section 4. In Section 5, we deal with stabilization of 
axion-dilaton modulus, using close string 3-form fluxes and 
show that we are able to stabilize all the close string moduli 
at a weak string coupling $g_s = \frac{1}{4}$. Some concluding
remarks are presented in Section 6, while Appendix A contains
technical details on $T^6$ parametrization that we mainly use
in Section 5.

\section{Magnetized $D9$-branes}\label{D9}

\subsection{Setup}\label{D9setup}

The stacks of $D9$-branes are characterized by three independent sets of 
data ; their multiplicities $N_a$, winding matrices $W_{a,\, I}^{\hat{I}}$ and 
first Chern numbers $m^a_{\hat{I} \hat{J}}$ of the $U(1)$ background on their  
world-volume $\Sigma_9^a$,  $a=1,\dots,K$ . The first describes the rank of the the 
unitary gauge group  $U(N_a)$ on each $D9$ stack. The second is the covering of the 
world-volume of each stack of $D9$-branes on the ambient space. In other words, it gives 
the winding of the branes around the different cycles of the internal space\cite{Bianchi:2005yz}. 
Their entries are therefore integrally quantized. The last set of parameters is the first 
Chern numbers of the $U(1)\subset U(N_a)$ background on the world-volume of the 
$D9$-branes. For each stack, a linear combination of the $U(N_a)$ generators lying in the 
Cartan subalgebra is chosen. It forms a $U(1)$ subalgebra whose constant field strength 
is introduced on the covering of the internal space. These are subject to the Dirac 
quantization condition. On the world-volume of each stack of $D9$-branes, they are therefore 
integrally quantized. 

In type I string theory, the number of magnetized $D9$-branes must be doubled. Indeed 
the orientifold projection ${\cal O}=\Omega_p$ is defined by the world-sheet parity, 
it maps the field strength $F_a =  dA_a$ of the $U(1)_a$ gauge potential $A_a$ to its inverse 
${\cal O}: F_a \rightarrow -F_a$. The magnetized $D9$-branes are therefore not an invariant 
configuration as it stands. For each stack, a mirror stack must be added with an inverse 
flux on their world-volume. The complete gauge group of this construction remains a 
product of unitary groups $\otimes_a U(n_a)$, since the associated open strings  attached 
on a given stack  are identified with the ones attached on the mirror stack. In addition 
to these vectors, the massless spectrum contains adjoint scalars and fermions which form 
a ${\cal N}=4$, $d=4$ supermultiplet. On the other hand,  open strings stretched between 
the $a$-th and $b$-th stack give rise to chiral spinors in the bifundamental representation 
$(N_a,\bar{N}_b)$. Their multiplicity $I_{ab}$ is given by the index theorem of the product 
bundle ${\cal E}_a \otimes {\cal E}_b$ associated to the $U(1)_a \times U(1)_b$ flux 
\cite{Bianchi:2005yz}.
\be
I_{ab} = C_3 ({\cal E}_a \otimes {\cal E}_b) = 
{ {\rm det}W_a{\rm det}W_b \over (2\pi)^3} \int_{T^6}\left( q_aF_a +q_bF_b\right)^3\, ,
\label{intersection}
\ee
where $F_{a}$ is the pullback of the integrally quantized world-volume flux 
$m^a_{\hat{I} \hat{J}}$ on the target torus, and $q_a$ the corresponding $U(1)_a$ charge;
in our case $q_a=+1$ (-1) for the fundamental 
(anti-fundamental representation). The transformation under the gauge group 
and their multiplicities are thus determined in terms of the data 
$(N_a, W_{a,\, I}^{\hat{I}},m_{\hat{I} \hat{J}})$. 

Open strings stretched between the $a$-th brane and its mirror $a^\star$ give rise 
to massless modes associated to $I_{aa^\star}$ chiral fermions. These transform either 
in the antisymmetric or symmetric representation of $U(N_a)$. In addition to the massless 
chiral spinors, there exist in all twisted open string sectors a set of massive  
(or tachyonic) scalars in the same representation as the associated spinor corresponding 
to their superpartners under the supersymmetries generally broken by the magnetic fluxes.

Let us be more specific and assume the presence of K stacks of $N_a$ magnetized 
$D9$-branes, $a=1, \dots, K $. Each  stack is associated with a corresponding $U(N_a)$ 
gauge symmetry.  We choose K linear combinations of the generators of $U(N_a)$ which 
lie in the Cartan subalgebra and denote their abelian gauge potentials by $A^a$; for
simplicity, we identify them with $U(1)_a$. Their  
field strengths  is assumed to take constant values on the torus $T^6$. Thus there is   
a set of $K$ $U(1)$ gauge potentials $A^a$ with constant background field strengths
\be
\label{stab:gen:A}
A^a_\alpha = {1\over 2} F^a_{\alpha\beta}X^\beta  \quad {\rm where} \quad a=1,\dots, K \, .
\ee
Moreover the magnetized $D9$-branes couple only to the $U(1)$ flux associated 
with the gauge fields located on their own world-volume. In other words, the charges  of 
the endpoints $q_R$ and $q_L $ of the open strings stretched between the $i$-th and the 
$j$-th $D9$ brane can be written as $q_L  \equiv q_i$ and $q_R \equiv -q_j$ and the Cartan 
generator $h$ is given by
$h = {\rm diag}(h_1\id_{N_1}, \dots , h_N\id_{N_K})$, with 
$\id_{N_a}$ being the $N_a\times N_a$ identity matrix.

The magnetized $D9$-branes are also characterized by their wrapping numbers around 
the different 1-cycles of the torus which are encoded in the covering matrices 
$W^{\hat{\alpha},\; a}_{\; \alpha}$ defined as
\be
\label{md9:W}
W^{\hat{I}}_{\; J} = {\partial \xi^{\hat{I}} 
\over \partial X^J} \quad \quad {\rm for} \;\;  {\hat{I}}, J=0,\dots ,9\, ,
\ee
where the coordinates on the world-volume are denoted by $\xi^{\hat{I}}$ 
while the coordinates on the spacetime ${\cal M}_{10}$ are $X^I$. Similarly to 
the space-time which is assumed to be the direct product of a four-dimensional 
Minkowski manifold with a six-dimensional torus,  the form of the covering matrix 
is assumed to be as
 \be
\label{md9:W2}
W_{a, \, J}^{\hat{I}} = 
\left(
\begin{array}{cc}
\delta_{\mu}^{\hat{\mu}} & 0 \\
0 & W_{\alpha}^{\hat\alpha\, , a}
\end{array}
\right)
\quad \quad {\rm for} \;\; \mu, \hat{\mu} = 0,\dots , 3 \;\; {\rm and } \;\; 
\alpha , \hat{\alpha} = 1,\dots , 6\, , 
\ee
where the upper block corresponds to the covering of the $\Sigma_4$ on the four-dimensional 
spacetime ${\cal M}_4$. Since it is assumed that both are identical, the associated 
covering map $W^{\hat{\mu}}_\mu$ is the identity, 
$W^{\hat{\mu}}_\mu = \delta^{\hat{\mu}}_\mu$. The entries of the lower block, 
on the other hand, describe the wrapping numbers of the $D9$ around the different 
1-cycles of the torus. These are therefore restricted to be integers 
$W^{\hat{\alpha}}_\alpha \in \mathbb{Z}$,  $\forall\;  \alpha, \hat{\alpha} = 1,\dots, 6$. 
 
The $K$ $D9$ stacks are then ten-dimensional objects which fill the four-dimensional 
space-time and cover the internal torus $T^6$. Thus there are $K $ different 
coverings ${\cal T}_6^a$ of the torus $T^6$ described by the $K$ covering maps 
$W^{\hat{\alpha},\; a}_{\; \alpha}$, for $a=1,\dots, K$. The fields $F^a_{\alpha\beta}$ 
then correspond to a non-trivial $U(1)$ gauge bundle on the torus $T^6$. Equivalently, 
their world-volume field strengths  $F^a_{\hat{\alpha}\hat{\beta}}$ correspond to a 
non-trivial  $U(1)$ gauge bundle on the covering ${\cal T}_6^a$ of the torus $T^6$.  
The Dirac quantization condition applies independently to the $K$ fluxes 
$F^a_{\hat{\alpha}\hat{\beta}}$
\be
\label{stab:gen:F}
\left\{
\begin{array}{ccc}
 F^a_{\hat{\alpha}\hat{\beta}} = m^a_{\hat{\alpha}\hat{\beta}} \in \;  
\mathbb{Z} &, \; \forall\hat{\alpha},\hat{\beta} = 1,\dots, 6&
\\
&&\quad \quad \forall a = 1,\dots, K
\\
p^a_{\alpha\beta} =  (W^{-1})_{\alpha}^{\hat{\alpha}, \; 
a} (W^{-1})_\beta^{\hat{\beta} , \; a} \,  
m^a_{\hat{\alpha}\hat{\beta}} \in \mathbb{Q} &, \; \forall \alpha,\beta = 1,\dots, 6&
\end{array}
\right.
\ee 

Note that these rationally quantized fluxes are equivalent to the one 
introduced in \cite{AM}. Here, the entries of the winding matrix describe 
the 1-cycle winding numbers, whereas in \cite{AM}, the space-time fluxes are 
defined via the 2-cycle winding numbers $n_{\alpha\beta}$. For the simplest flux 
configurations, the winding numbers $n_{\alpha\beta}$ are the denominators of the 
entries of the matrix $p^a_{\alpha\beta}$.

\subsection{Consistency conditions}\label{D9cons}

Necessary conditions for a consistent construction involving $K$ stacks of 
$N_a$  magnetic $D9$-branes on a compact orientifold compactification follow from
the Ramond-Ramond (R-R) tadpole cancellations. These account for the absence 
of UV divergencies in the one loop amplitude and ensure, via a generalized 
Green-Schwarz mechanism, the cancellation of gauge anomalies in the associated 
four dimensional field theories. In the toroidal compactification of type I string 
theory, the magnetized $D9$-branes induce 9-brane and 5-brane charges, while the 
3-brane and 7-brane charges automatically vanish due to the presence of mirror 
branes with opposite flux. For general magnetic fluxes, they can be written in 
terms of the Chern numbers and winding matrix \cite{Bianchi:2005yz}. 
The tadpole conditions read in this case
\bea
16 &=& \sum_{a=1}^K \; N_a\;  {\rm det}W_a
\label{tad9}
\\
0  &=& \sum_{a=1}^K\; N_a \; {\rm det}W_a \; 
{\cal Q}_a^{\alpha\beta} \qquad \forall \alpha,\beta=1,\dots,6
\label{tad5}
\eea
where
\be
{\cal Q}_a^{\alpha\beta} = \epsilon^{\alpha\beta\delta\gamma\sigma\tau} 
p^a_{\delta\gamma}p^a_{\sigma\tau} \, .
\label{Qdef}
\ee
The l.h.s of eq.~(\ref{tad9}) arises from the charge contribution of the 
$O9$ plane. Moreover,  the toroidal compactification implies the absence of 
any $O5$-planes and thus  the l.h.s of eq.~(\ref{tad5}) vanishes. 
Note that equations (\ref{tad5}) are in agreement with the 5-brane tadpole condition 
given in \cite{AM}, where the factor ${\cal K}$ accounts for the $\epsilon$ tensor in
eq.~(\ref{Qdef}), while the 9-brane tadpole condition (\ref{tad9}) disagrees with 
\cite{AM} because the factor ${\cal K}$ should be absent.\footnote{We thank
F. Denef and F. Marchesano for correspondence and enlightening discussions
on this point.}

The above tadpole conditions restrict the allowed choices for the ranks of 
the gauge groups, winding matrices, Chern numbers and consequently the allowed 
spectra. They ensure in particular that the spectrum is anomaly-free. Note 
also that they are invariant under a discrete  rescaling of the Chern numbers 
in some given direction for all stacks, keeping the winding matrix invariant
\be
\{m^a_{\hat{\alpha}\hat{\beta}},W_{a,\, \alpha}^{\hat{\beta}}\} 
\rightarrow \{\Lambda \, m^a_{\hat{\alpha}\hat{\beta}},
W_{a,\, \alpha}^{\hat{\beta}}\}\qquad \forall a=1,\dots, 
K \quad {\rm and} \quad {\rm for \; a \; given\; } 
{\hat{\alpha}}\quad ; \quad \Lambda \in \mathbb{Z}. 
\label{rescaling}
\ee
This rescaling affects the spectrum at the intersection of all pairs of stacks. 
Form eq.~(\ref{intersection}), we see that the number of chiral fermions in all 
intersections is also rescaled. Note that this invariance of the tadpole 
conditions is not true anymore in the case of orbifold compactifications 
with $O5$-planes. In this case, the $O5$-planes carry a R-R charge which is 
not insensitive to the rescaling (\ref{rescaling}). Since these contribute to 
the 5-brane tadpole (\ref{tad5}), the consistency conditions would not be invariant. 

\subsection{Supersymmetry conditions}\label{D9susy}

For a given configuration of magnetized branes, one may ask whether the 
different stacks forming the brane configuration preserve some common supersymmetries.  
Via a Zeeman-like effect,  the scalars and fermions of the twisted sectors, 
corresponding to the intersection of two brane stacks acquire different masses. 
Supersymmetry is thus broken \cite{Bachas:1995ik}. Similarly, a single magnetized 
$D9$-brane in  type I string theory is not generically supersymmetric. 
Indeed, the orientifold projection implies the presence of mirror 
branes. Twisted scalars from the Neveu-Schwarz (NS) sector of open string 
stretched between a brane and its image are generically massive, while some 
chiral spinors from the Ramond (R) sector remain massless. In other words, 
the $D9$-brane does not preserve the same supersymmetry as the orientifold 
projection. Part of supersymmetry may however be restored. The supersymmetry 
conditions involve the flux quanta and winding matrix, but also the metric moduli.

 In the case of toroidal compactification of type I string theory, starting 
from the real orthonormal basis of $T^6$: $x_i = x_i + 1$ and $y^i = y^i + 1$, 
$i=1,2,3$ with unit periodicity, the moduli decompose in a complex structure 
variation which is parametrized by the matrix $\tau_{ij}$ entering in the 
definition of the complex coordinates $z_i = x_i + \tau_{ij}y^j$  and 
in the variation of the mixed part of the metric  described by the 
real $(1,1)$-form $J = i\delta g_{i\bar{j}} dz^i \wedge d\bar{z}^j$. 
The supersymmetry conditions then read\cite{AM}
\be
F_{(2,0)}^a = 0\ \ ; \quad {\cal F}_a \wedge {\cal F}_a \wedge {\cal F}_a 
= {\cal F}_a \wedge J \wedge J\ \ ; \quad {\rm det}W_a 
\left(J \wedge J \wedge J - {\cal F}_a \wedge {\cal F}_a \wedge J\right)>
0 \quad \forall a = 1,\dots, K \, .
\label{susy}
\ee
The complexified fluxes can be written as
\bea
F^a_{(2,0)} & = & {(\tau-\bar{\tau})^{-1}}^T \! \left[ \tau^{T} p^{a}_{xx} 
\tau - \tau^{T}{p^{a}_{xy}} - p^{a}_{yx}\tau + 
p^{a}_{yy}\right] (\tau-\bar{\tau})^{-1} \label{stab:gen:purelyholo2}\\
F^a_{(1,1)} & = & {(\tau-\bar{\tau})^{-1}}^T\! \left[ -\tau^{T} 
p^{a}_{xx}\bar{\tau} + \tau^{T}{p^{a}_{xy}} + p^{a}_{yx}\bar{\tau} - 
p^{a}_{yy}\right] (\tau-\bar{\tau})^{-1}
\eea
where the matrices $(p^{a}_{xx})_{mn}$, $(p^{a}_{xy})_{mn}$ and 
$(p^{a}_{yy})_{mn}$
enter in the quantized field strength (\ref{stab:gen:F}) in the directions
$(x^m,x^n)$, $(x^m,y^n)$ and $(y^m,y^n)$, respectively. The field strengths
 $F^a_{(2,0)}$ and $F^a_{(1,1)}$ are $3\times 3$
matrices that correspond to the upper half of the matrix 
$\mathcal{F}^a$:
\be
\mathcal{F}^a=-(2\pi)^2 i\alpha' \left(
\begin{array}{cc}
F^a_{(2,0)} & F^a_{(1,1)}\\
-{F^a}^\dagger_{(1,1)} & {{F^a}^*}_{(2,0)}\\
\end{array}
\right)\, ,
\label{matrixF}
\ee
which is the total field strength in the 
cohomology basis $e_{i\bar{j}} = i dz^i \wedge dz^j$.

The first set of conditions of eq.~(\ref{susy}) states that the purely holomorphic 
flux vanishes. For given flux quanta and winding numbers, this matrix equation 
restricts the complex structure $\tau$. Using eq.~(\ref{stab:gen:purelyholo2}), 
the supersymmetry conditions for each stack can first be seen as a restriction on the 
parameters of the complex structure matrix elements $\tau$:
\be
F_{(2,0)}^a = 0\qquad  \rightarrow \qquad \tau^{T} p^{a}_{xx} \tau - 
\tau^{T}{p^{a}_{xy}} - p^{a}_{yx}\tau + 
p^{a}_{yy} = 0\, .
\label{stab:gen:M20_condition}
\ee
Similarly with the tadpole conditions,  the ``holomorphicity" 
equation (\ref{stab:gen:M20_condition}) is invariant under a general rescaling of all fluxes 
\be
m^a_{\hat{\alpha}\hat{\beta}} \rightarrow \Lambda^a \; 
m_{\hat{\alpha}\hat{\beta}}^a \qquad ; \qquad \forall\; 
\hat{\alpha},\hat{\beta}=1,\dots,6\qquad {\rm and} \; a=1,\dots, K \; .
\label{rescaling2}
\ee
This may be compared to the tadpole conditions which have a wider invariance under 
rescaling of Chern numbers in given direction, while the conditions 
(\ref{stab:gen:M20_condition}) are only invariant under the same rescaling 
in all directions.   

The second set of conditions  of eq.~(\ref{susy}) gives rise to a real 
equation and restricts the K\"ahler moduli. This can be understood as a 
$D$-flatness condition. In the four-dimensional effective action, the 
magnetic fluxes give rise to  topological couplings for the different axions of 
the compactified field theory. These arise from the dimensional reduction of the 
Wess Zumino (WZ) action. In addition to the topological coupling, the ${\cal N}=1$ 
supersymmetric action yields a Fayet-Iliopoulos (FI) term of the form 
\be
{\xi_a\over g_a^2} = {1\over (4\pi^2 \alpha')^3}
\int_{T^6}\big( {\cal F}_a \wedge {\cal F}_a\wedge {\cal F}_a -
{\cal F}_a\wedge J \wedge J\big)\, .
\ee
The $D$-flatness condition in the  absence of charged scalars requires then 
that $<D_a> = <\xi_a> = 0$, which is equivalent to the second equation of eq.~(\ref{susy}). Finally, the last inequality  in eq.~(\ref{susy}) may also be understood from a four-dimensional viewpoint  as the positivity of the $U(1)_a$ gauge coupling $ g_a^2$, since its  expression  in terms of the fluxes and moduli reads
\be
{1\over g_a^2} = {1\over (4\pi^2 \alpha')^3} \int_{T^6}\big( J \wedge J\wedge J -{\cal F}_a\wedge {\cal F}_a \wedge J\big)\, .
\label{gaugecoupling}
\ee

The above supersymmetry conditions are only valid in the absence of charged scalars. The situation in the presence of scalars charged under the $U(1)$ gauge groups is different. The $D$-flatness condition is modified. In the low energy field theoretical approximation, the D-term reads 
\be
D_a = - \left( \sum q^\phi_a |\phi_a|^2 + M_s^2\xi_a \right) \, ,
\label{dterm}
\ee
where $M_s=\alpha'^{-1/2}$ is the string scale\footnote{When mass scales are absent, string units are implicit throughout the paper.}, and the sum is extended over all
scalars $\phi_a$ charged under the $a$-th $U(1)_a$ with charge $q_a^{\phi}$. Such scalars arise in the compactification of magnetized $D9$-branes in type I string theory, for instance from the NS sector of open strings stretched between the $a$-th brane and its image $a^\star$. When one of these scalars acquire a non-vanishing VEV $<|\phi_a|^2 > = v_a^2$,  the calibration condition of eq.~(\ref{susy}) is modified to 
\bea
q_a v_a^2 \int_{T^6}\big( J \wedge J\wedge J -{\cal F}_a\wedge {\cal F}_a \wedge J\big) &=& -M_s^2{\int_{T^6}\big( {\cal F}_a \wedge {\cal F}_a\wedge {\cal F}_a -{\cal F}_a\wedge J \wedge J\big)}
\label{stab:gen:cond_kahlerX}
\\
{\det W}_a\left(J \wedge J \wedge J - {\cal F}_a \wedge {\cal F}_a \wedge J \right)&>& 0 \quad , \, \quad \forall a=1,\dots, K \, .
\label{stab:gen:cond_pos}
\eea

In contrast with the "holomorphicity" equation (\ref{stab:gen:M20_condition}), the 
conditions (\ref{stab:gen:cond_kahlerX})  and (\ref{stab:gen:cond_pos}) are not invariant 
under the rescaling (\ref{rescaling2}) . This now leads to a family of solutions which 
differ by their overall volume. Indeed, the rescaling (\ref{rescaling2}) at fixed winding 
numbers corresponds to the rescaling of all fluxes ${\cal F}_a$. If all stacks scale 
in the same way $\Lambda_a \equiv \Lambda$, $\forall a =1,\dots, K$, one obtains an 
infinite family of solutions
\be
\{ {\cal F}_{a} , J \} \; \rightarrow \; \{ \Lambda \; {\cal F}_{a} , 
\Lambda \; J \} \quad  {\rm for} \qquad \Lambda \in \mathbb{N} \quad {\rm and } \; 
\quad \forall a =1,\dots, K\, .
\ee
Since the tadpole and holomorphicity conditions are invariant under this rescaling, 
one obtains an infinite discretum of  vacua which differ by their spectra and the overall 
volume. All of them have the same gauge symmetry but with different gauge 
couplings (\ref{gaugecoupling}). Similarly, their total internal volume and
consequently their four-dimensional Planck mass differs by a factor of $\Lambda^3$. 
However, not all of these vacua are phenomenologically viable. 
Indeed, the experimental bounds on the string scale, on the numbers of chiral 
families and on the value of the longitudinal volumes strongly restricts the permissible 
vacua. Nevertheless, all of them are fully consistent from the viewpoint of string theory. 

It turns out that there exist no toroidal supersymmetric models of magnetized $D9$-branes 
with chiral matter in the literature. Despite the absence of a  full no-go theorem, 
it is widely believed that the tadpole conditions (\ref{tad9}) and (\ref{tad5}) are 
not compatible with the supersymmetry conditions (\ref{susy}) at zero open string 
VEV's.\footnote{The examples found in \cite{AM} were due to the presence of the sign
factor $\cal K$ in the 9-brane tadpole condition (\ref{tad9}).} In the following section,
we show that this is not true when one turns on a non-vanishing VEV's for charged 
scalars on the branes. Actually,  one such VEV, which does not even break gauge 
symmetries, is sufficient to render compatible the supersymmetry with the tadpole
conditions.

\section{Supersymmetric toroidal model}\label{STM}

Our aim is first to show that the tadpole equations are compatible with the deformed 
supersymmetry condition (\ref{stab:gen:cond_kahlerX}). An explicit example of a 
consistent configuration  of supersymmetric $D9$-branes is presented where the tadpole 
conditions (\ref{tad9}) and (\ref{tad5}) are satisfied. We will give an example of nine 
magnetized branes which are supersymmetric for fixed values of the  metric moduli of the torus. 
The only remaining closed string modulus is then the dilaton and the associated axion field. 
One can then analyze the case of usual toroidal models in the limit of vanishing VEV's for the 
charged scalars. This will be shown to approach points at the boundary of moduli space 
corresponding to the decompactification limit where the volume of the internal torus 
becomes infinite. We will finally show the existence of many infinite discreta of vacua 
which differ by their spectra, gauge couplings and four-dimensional Planck mass in string units. 

To this end, one slightly modifies the configuration of branes presented in \cite{AM}. 
Inspection of eqs.~(\ref{stab:gen:M20_condition}) and (\ref{stab:gen:cond_kahlerX}) 
shows that for each stack of magnetized $D9$-branes, we have up to 
three complex conditions for the moduli of the complex structure, 
depending on the directions in which the fluxes are switched on, 
whereas only one real condition can be set on the K\"ahler moduli. 
Therefore, to fix all K\"ahler moduli in a toroidal compactification, 
at least nine stacks of branes must be added.
The first  six branes have \emph{oblique} fluxes on their world-volume. 
They do not have any chiral fermions on their intersections and preserve  
(each one) an ${\cal N}=2$, $d=4$ supersymmetry for restricted complex structure 
and K\"ahler moduli. Moreover in our example, since the number of
of intersections of any pair vanishes, the intersections preserve also
extended ${\cal N}=2$ supersymmetry, although not the same for each pair.
All complex structure moduli are fixed, while three K\"ahler 
moduli remain undetermined. These are stabilized, in terms of a single 
charged scalar VEV, by the three last stacks which have usual 
parallel fluxes. Their intersections are generically not trivial and the massless 
spectrum contains chiral ${\cal N}=1$ supermultiplets.

\begin{table} 
\vskip-0.5cm
\begin{center}
$$
\begin{array}{|c||c|c|c|c|}
\hline
\mathrm{Stack} \sharp & \mathrm{Multiplicity} &\mathrm{Fluxes} & \mathrm{Fixed\ moduli} & 
\mathrm{5-brane\ localization}\\
\hline
\hline
&&&&\\
\sharp 1  & N_1=1&(F^1_{x_1 y_2},F^1_{x_2 y_1})=(1,1)& 
\tau_{31} = \tau_{32}=0 & [x_3,y_3]\\ 
          &                          &     &  \tau_{11}= 
\tau_{22}  &           \\
 &                          &     & {\rm Re}J_{1\bar{2}}=0  &           \\
&&&&\\
\hline
 &&&&\\
\sharp 2  & N_2=1& (F^2_{x_1 y_3},F^2_{x_3 y_1})=(1,1)& 
\tau_{21} = \tau_{23}=0 &  [x_2,y_2]\\
          &                          &     & \tau_{11}= 
\tau_{33} & \\
 &                          &     & {\rm Re}J_{1\bar{3}}=0  &           \\
&&&&\\
\hline
 &&&&\\
\sharp 3  & N_3=1& (F^3_{x_1 x_2},F^3_{y_1 y_2})=(1,1)& 
\tau_{13}=0\, , \, {\rm Im}J_{1\bar{2}}=0& [x_3,y_3] \\
          &                          && 
\tau_{11}\tau_{22}=-1 & \\
&&&&\\
\hline 
\hline
&&&&\\
\sharp 4  & N_4=1& (F^4_{x_2 x_3},F^4_{y_2 y_3})=(1,1)& 
\tau_{12}=0 \, , \, {\rm Im}J_{2\bar{3}}=0  & [x_1,y_1] \\
&&&&\\
\hline
&&&&\\
\sharp 5 & N_5=1& (F^5_{x_1 x_3},F^5_{y_1 y_3})=(1,1)&  {\rm Im}J_{1\bar{3}}=0 & 
 [x_2,y_2]\\
&&&&\\
\hline
&&&&\\
\sharp 6  & N_6=1& (F^6_{x_2 y_3},F^6_{x_3 y_2})=(1,1)&  {\rm Re}J_{2\bar{3}}=0 & 
 [x_1,y_1]\\
&&&&\\
\hline
\end{array}
$$
\end{center}
\vskip-0.5cm
\caption{ Fixed complex structure moduli for each magnetized stack 
$\sharp$ of $ D9$-branes  depending on the quantized fluxes. The last 
column gives the localization on the 2-cycles  $[x_i,y_i]$, 
of the induced 5-brane charges.}
\label{table1}
\end{table}

\subsection{Explicit Model}\label{explicit}

Here, we prove the  existence of a family of supersymmetric toroidal compactifications with magnetized $D9$-branes. The presence of VEV's for the charged fields cures the apparent incompatibility between the supersymmetry condition (\ref{susy}) and the tadpole conditions (\ref{tad9}) and (\ref{tad5}). 

\begin{table}
\begin{center}
$$
\begin{array}{|c||c|c|c|}
\hline
\mathrm{Stack} \sharp & \mathrm{Multiplicity} &\mathrm{Fluxes} & D5 \mathrm{branes \, \, 
localization }\\
\hline\hline
&&&\\
\sharp 7  & N_7=1& (F^7_{x_1 y_1},F^7_{x_2 y_2},0)=(2,-3,0)  & [x_3,y_3]
\\ 
          &     &                    &\\
\hline
 &&&\\
\sharp 8  & N_8=3& (F^8_{x_1 y_1},0,F^8_{x_3 y_3})=(-2,0,1)  & [x_2,y_2]
  \\
          && &\\
\hline
 &&& [x_1,y_1]  \\
\sharp 9  & N_9=2& (F^9_{x_1 y_1},F^9_{x_2 
y_2},F^9_{x_3 y_3})=(4,1,1)&  [x_2,y_2] \\
          &                          && [x_3,y_3]\\\hline 
\end{array}
$$
\end{center}
\caption{Additional stacks of magnetized $D9$-branes allowing the 
stabilization of the diagonal part of the K\"ahler form. The last 
column gives the localization on the 2-cycles $[x_i,y_i]$, of the induced 5-brane charges.}
\label{table2}
\end{table}

The model is then constructed out of the nine stacks presented in Tables \ref{table1} 
and \ref{table2}, with all winding matrices $W_a$ equal to the identity. 
Following the setup of \cite{AM}, the first six branes have purely 
oblique fluxes and each preserves separately  ${\cal N}=2$ supersymmetry. They fix the 
complex structure moduli to be of the form
\be
\tau_{ij} = i\delta_{ij} \, ,
\label{tausol}
\ee
and all off-diagonal components of the K\"ahler form  to  be vanishing, 
\be
J_{i\bar{j}} = 0\, .
\label{Joffsol}
\ee
This geometry corresponds to factorizable tori 
$T^6 = T^2 \times T^2 \times T^2$, where each $T^2$ is a squared lattice. 
Furthermore, the contribution of these off-diagonal fluxes to the 5-brane 
tadpoles is diagonal and negative. It sums up to
\be
(Q_5^{x_1y_1},Q_5^{x_2y_2},Q_5^{x_3y_3}) = (-2,-2,-2)\, .
\label{tad1-6}
\ee

The last three brane stacks stabilize the remaining three K\"ahler moduli $J_{i\bar{i}}$. 
The minimal modification of the setup of \cite{AM} is the addition of a non-vanishing 
VEV for a single scalar field charged under one of the last  three $U(1)$'s, for instance 
the ninth, $v^2_9 \neq 0$. In fact, the choice of quanta of Table \ref{table2} cancels 
the tadpole (\ref{tad1-6}) induced by the first six stacks. Furthermore, these are 
supersymmetric for restricted values of the K\"ahler moduli. Indeed, from the conditions 
(\ref{stab:gen:cond_kahlerX}), these  are fixed to be 
\be
2J_{3\bar{3}} = {2 \over 3} J_{2\bar{2}}=J_{1\bar{1}} :=  J\, ,
\ee
where
\be
q{v_9^2 \over M_s^2}= -\xi_9=-{ 16 - 20J^2 \over 3 J^3 - 36 J^2}\qquad {\rm with}\quad q=2\, .
\label{solK}
\ee
The VEV $v_9$ corresponds to a charged scalar in the antisymmetric representation
of $U(N_9)$ with $U(1)_9$ charge $q=2$; there are no states in the symmetric
representation since the winding number is 1.
The relative sign between the value of $q$ and the FI parameter $\xi_9$, appearing
in the $D$-term (\ref{dterm}), can be easily verified by the presence of a tachyonic
state in the spectrum in the large volume limit, according to the formula (\ref{solK}).
For the above particular points of the K\"ahler moduli space, the $D$-flatness 
condition (\ref{stab:gen:cond_kahlerX}) is satisfied, while keeping the gauge 
coupling positive, or equivalently satisfying the condition (\ref{stab:gen:cond_pos}).
Finally, since each brane contributes one unit of 9-brane R-R charge in the r.h.s. 
of eq.~(\ref{tad9}), the total contribution is 
\be
Q_9=12\, .
\ee
One can then add for instance four
extra non-magnetized $D9$-branes to account for the left-over charge.

Note that our computation is valid for small values
of $v_9$ (in string units), since the inclusion of the charged scalars in the
$D$-term is in principle valid perturbatively. From eq.~(\ref{solK}), this corresponds
to large values of the K\"ahler parameter $J$. Actually, this equation is valid for
$J>12$ in order to satisfy the positivity condition (\ref{stab:gen:cond_pos}).
It follows that in this region there is always a solution of the supersymmetry
equation (\ref{solK}). Consequently, the above
configuration of nine magnetized $D9$-branes forms a consistent supersymmetric 
model where all metric moduli are fixed. Moreover from the usual St\"uckelberg 
couplings, the R-R moduli are absorbed in the longitudinal polarization of the 
$U(1)$ gauge bosons. All nine $U(1)$ gauge fields become then massive. The gauge 
symmetry of this vacuum  is therefore
\be
SU(3)\times SU(2) \times SO(8)
\ee
where the last factor corresponds to the four additional non-magnetized $D9$-branes 
needed to satisfy the 9-brane tadpole condition (\ref{tad9}).
To obtain the standard model gauge group, an extra $U(1)$ factor  can easily be added in a supersymmetric configuration, reducing the $SO(8)$ symmetry to $SO(6)$. 
The St\"uckelberg couplings give mass to 
nine $U(1)$ gauge bosons, while in general a linear combination of the ten $U(1)$'s
remains massless. One can therefore obtain an additional abelian factor with chiral spectrum. 

Finally, one may expect that the presence of non-trivial VEV's for charged scalars break
the gauge symmetry. This is not in general true. For instance in our example, as we saw
above, the charged scalars from the ninth stack transform in the antisymmetric representation
of $U(2)$ which is $SU(2)$ singlet. Moreover, due to the St\"uckelberg coupling, the $U(1)$ 
gauge boson is massive and the abelian gauge symmetry is already broken.
Nevertheless, the left-over global $U(1)$ symmetry is spontaneously broken by the 
presence of $v_9$, signaled by the presence of a Goldstone boson. In the presence of an additional magnetized D9-brane stack that stabilizes the VEV of the charged scalar, the above Goldstone boson is absorbed in the new associated $U(1)$ gauge field that becomes massive by the usual Higgs mechanism. This will become clear in an explicit example that we present below, in Section \ref{openstab}.

\subsection{Extensions to other models}\label{D9extension}

The non-vanishing VEV $v_9$ appearing in eq.~(\ref{solK}) corresponds to charged 
scalars from the NS sector of open strings stretched between the ninth brane stack 
and its mirror. At the supersymmetric points $<D>=0$, a linear combination of the 
K\"ahler form and the open string scalar remains massless. Thus, in this example,
the direction of the charged field is flat and there 
exist no preferred values for its VEV $v_9$. In Section \ref{openstab} below, we show how this VEV can be fixed by adding
extra stacks of magnetized branes. Indeed, we present an explicit example, where
the last three stacks $\sharp 7,8,9$ of the model described in Section \ref{explicit} are
replaced by five others with the following properties: (1) Three of them are used
to fix the three
diagonal K\"ahler moduli for vanishing scalar VEV's, in terms of the magnetic
fluxes. (2) The remaining two are used to satisfy tadpole cancellation conditions,
while supersymmetry requirement fix two charged scalar VEV's to be non-vanishing,
at values smaller than the string scale consistently with the $\alpha'$-expansion.
Actually, even in the example presented above as we mentioned already,
the field theoretical approximation which led to the 
$D$-flatness condition (\ref{stab:gen:cond_kahlerX}) is only valid for 
small VEV $v_9^2 \ll M_s^2$, since higher powers in $|\phi|^2$ have been neglected. 
This yields minima at large values for the K\"ahler moduli $J\gg \alpha'$. 

 Note also that the usual compactification in the absence of VEV's for the charged scalars 
can be obtained in the limit where $v_9 \rightarrow 0$. From eq.~(\ref{solK}), one 
observes that it corresponds to the decompactification limit where the overall volume 
of the torus becomes infinite, $J \wedge J \wedge J \rightarrow \infty$. In other words 
traditional toroidal compactification of magnetic branes may lead to supersymmetric 
vacua only at the boundary of the moduli space.

Further models can finally be constructed apart from the one presented above. Moreover,
aside from the discretum of vacua arising by the general rescaling of the fluxes and volume, 
there exists a second one. It is characterized by a discrete set of volumes of one of 
the three $T^2$, while keeping the other volumes, as well as the shape moduli, invariant. 
In other words, this family of vacua is given by the set of volumes
\be
J^\Lambda = (\Lambda^2 J,\ {3\over 2}J,\ {1\over 2}J)
\label{solLambda}
\ee
where $J$ is always a solution to the equation (\ref{solK}). For this, the 
Chern numbers $m^b_{x_1y_1}$ of the last three stacks $b=7,8,9$ are rescaled 
by $\Lambda^2$, $m^b_{x_1y_1} \rightarrow \Lambda^2 \; m^b_{x_1y_1}$. 
This modifies also the 5-brane tadpole charges induced by the last three stacks:
\be
Q_5 = (2,\, 2\Lambda^2,\, 2\Lambda^2)\, .
\ee
In order to compensate them in such a way that the complex structure remains of the 
diagonal form (\ref{tausol}), the fluxes of the stacks $\sharp 1$, $\sharp 2 $, $\sharp 3$ 
and $\sharp 5$ must be rescaled such that  
\be
m^{1,\Lambda}_{x_1y_2}=
\,\,\,\,\,m^{2,\Lambda}_{x_1y_3}=
 m^{3,\Lambda}_{x_1x_2} = m^{5,\Lambda}_{x_1x_3} = \Lambda\, .
 \ee
The tadpole conditions are then satisfied and the rescaled model is consistent. The
metric remains diagonal describing the product of three orthogonal $T^2$'s, while 
the volume of the first $T^2$ is rescaled by $\Lambda^2$ and the other two remain
the same. The replication of chiral fermions, the gauge couplings and the Planck 
mass are also affected since the total internal volume is rescaled.
 
 By a similar argument one can show that it is possible to rescale the areas of two
 of the tori, while keeping the third one fixed, leading yet to another discretum of vacua.
 
 \subsection{Hierarchy Problem}\label{HP}
 
From the analysis of Sections \ref{D9susy} and \ref{D9extension}, one observes the 
possibility to obtain large volume compactifications \cite{ld} having two, four or six large dimensions. 
Here, we study the consequence on the hierarchy of the string scale with respect to the Planck mass.

If one assumes that the standard model lives on some of the magnetized $D9$-branes, 
the dimensions longitudinal to these branes are constrained by accelerator experiments. 
They can not be hierarchically larger than the string scale.  On other hand, since the 
$D9$-branes are space-time filling, there exist no dimensions transverse to their 
world-volume. The moduli stabilization at very large volume can not therefore be 
implemented to decouple the string scale from the Planck mass.

However, the model presented in Section \ref{explicit} can be easily modified 
to allow the presence of unmagnetized $D5$-branes, transverse for instance 
to the first torus with large value. In order for them to preserve the same 
supersymmetry as the magnetic fluxes, their GSO projection must be chosen 
in a way that their tension and R-R charge have opposite sign. Moreover, in the sectors of 
open strings stretched between magnetized $D9$-branes and the 
unmagnetized $D5$-branes, there exist chiral fermions. 

\subsection{Charged scalar VEV's determination}\label{openstab}

In the construction presented in section \ref{explicit}, the VEV $v_9$ of the charged field is
undetermined, corresponding to an open string modulus with flat potential. Here, we present an
extension based on the same principle of magnetized branes, where such open string scalar 
VEV's are fixed. To this end, one introduces more than nine stacks of magnetized $D9$-branes. 
The vanishing of the extra $D$-terms induces further conditions. Once the nine metric moduli are stabilized, the new conditions may then stabilize the above open string moduli that enter the $D$-terms. 

Let us present an explicit example, where besides the metric moduli we turn on VEV's for two 
massless charged fields. The latter are charged under two additional $U(1)$'s. These are 
embedded in a model defined by eleven  stacks of magnetized branes. The first six are those with oblique flexes given in Table \ref{table1}. The next five branes are new and given in 
Table \ref{tableV2}. The $D$-flatness condititions for the first nine stacks restrict the metric moduli of the torus to be of the diagonal form
\be
\tau_{ij} = i\delta_{ij} \quad \quad ; \quad \quad 
(J_{x_1y_1},J_{x_2y_2},J_{x_3y_3}) = 4\pi^2 \alpha' \sqrt{3 \over 22}(44,66,19) \, ,
\label{V:J+T}
\ee
in the absence of VEV's for the fields charged under these nine branes. The tadpole conditions (\ref{tad9}) and (\ref{tad5}) ask however for additional branes. These are the stacks $\sharp 10$ and 
$\sharp 11$, as well as four unmagnetized $D9$-branes. Due to the usual St\"uckelberg 
couplings, this model defines a consistent brane configuration with gauge symmetry
\be
 SU(3) \times SU(2)^3\times U(1)^2\, .
\ee
However, it can be supersymmetric only in the presence of non-trivial VEV's for open string states charged under the $U(1)$ gauge bosons of the last two magnetized $D9$-brane stacks. 

Let us then switch on VEV's for the fields $\phi_{10}$ and $\phi_{11}$, transforming in the
antisymmetric representations of the corresponding $SU(2)$ gauge groups and charged 
under the $U(1)$'s of the last two stacks. Their respective VEVs $v_{10}$ and  $v_{11}$, 
are fixed by the supersymmetry conditions. Indeed, from the quanta given in 
Table \ref{tableV2} and the values for the K\"ahler moduli (\ref{V:J+T}), the positivity 
conditions (\ref{stab:gen:cond_pos}) for these branes are satisfied. Moreover, since the 
K\"ahler form is already fixed, the supersymmetry conditions (\ref{stab:gen:cond_kahlerX}) 
determine the values of $v_{10}$ and  $v_{11}$ as:
\be
v_{10}^2 l_s^2  \simeq {0.71\over q}\simeq 0.35 \quad \quad ; 
\quad \quad  v_{11}^2 l_s^2  \simeq {0.31\over q}\simeq 0.15\, ,
\label{V:V}
\ee
where we used that the $U(1)$ charge of the fields in the antisymmetric representation is $q=2$
in our conventions, as mentioned earlier. 
These VEV's break the two $U(1)$ factors and the gauge group becomes 
$SU(3) \times SU(2)^3$. Note that the above values of the VEV's are reasonably small
in string units, consistently with our perturbative approach of including the charged scalar
fields in the $D$-terms.
We have thus presented a model where the open string moduli corresponding to charged
scalar VEV's are also fixed by the magnetic fluxes. In principle, the same method can be
applied for stabilizing other open string moduli, as well.

\begin{table}
\begin{center}
$$
\begin{array}{|c||c|c|}
\hline
\mathrm{Stack} \sharp & \mathrm{Multiplicity} &\mathrm{Fluxes} \\
\hline\hline
&&\\
\sharp 7  & N_7=1& (F^7_{x_1 y_1},F^7_{x_2 y_2},0)=(-4,-4,3)  
\\ 
          &     &                    \\
\hline
 &&\\
\sharp 8  & N_8=2& (F^8_{x_1 y_1},0,F^8_{x_3 y_3})=(-3,1,1)  
  \\
          && \\
\hline
 && \\
\sharp 9  & N_9=3& (F^9_{x_1 y_1},F^9_{x_2 
y_2},F^9_{x_3 y_3})=(-2,3,0) \\
          &                          &\\
          \hline 
 && \\
\sharp 10  & N_{10}=2& (F^{10}_{x_1 y_1},F^{10}_{x_2 
y_2},F^{10}_{x_3 y_3})=(5,1,2) \\
          &                          &\\
    \hline 
 && \\
\sharp 11  & N_{11}=2& (F^{11}_{x_1 y_1},F^{11}_{x_2 
y_2},F^{11}_{x_3 y_3})=(0,4,1) \\
          &                          &\\
          \hline
            \end{array}
$$
\end{center}
\caption{Additional stacks of magnetized $D9$-branes allowing the 
stabilization of the diagonal part of the K\"ahler form and some charged moduli.
}
\label{tableV2}
\end{table}

\section{Magnetized $D7$-branes}\label{D7}

\subsection{Generalities}

In previous sections, we have shown that  all the complex structure and K\"ahler class
moduli are stabilized using magnetic fluxes on $D9$-brane world-volume, when 
charged scalars acquire VEV's. However, one is 
still left with an unstabilized axion-dilaton modulus. A mechanism to
implement this stabilization is to use closed string 3-form fluxes \cite{Kachru:2002he}.
In this section, we show how the two kinds of fluxes, namely the magnetic and 
3-form fluxes, can be simultaneously turned on in a consistent way in order to 
stabilize all closed string moduli. To this end,
it has to be imposed that the supersymmetry preserved by the 3-form fluxes
is the same as the one preserved by an $O3$ plane,
implying that these fluxes $G_{(3)} \equiv H_{(3)} - \tau F_{(3)}$ 
are primitive and of the type $(2, 1)$:
\begin{equation}
G_{(3)} \wedge J =0 \quad ; \quad {G_{(3)}} \in H^{2,1}\, ,
\end{equation}
where $H_{(3)}$ and $F_{(3)}$ are the field strengths of the NS-NS and R-R 2-forms.
In this case, the compactification is therefore chosen to be the torus orientifold 
$T^6/\mathbb{Z}_2$, where the $\mathbb{Z}_2$ orientifold action 
${\cal O}_2 = \Omega_p G_2(-1)^{F_L}$ is a composition of the world-sheet 
parity $\Omega_p$ with the left fermion number $(-1)^{F_L}$ and the orbifold 
$G_2$ acting on the torus coordinates $G_2: z_i \rightarrow -z_i$, $\forall i=1,2,3$.
This induces  $O3$ planes that carry an overall charge $Q_{O3} = -16$.

In this closed string background, one may introduce $Dp$-branes. The presence of 
$Dp$-branes is however constrained by the 
Freed-Witten (FW) anomaly \cite{Freed:1999vc,Maldacena:2001xj}.  
NS-NS 3-form flux $H_{(3)}$ implies that the Bianchi identity 
for the gauge-invariant world-volume gauge field strengths ${\cal F}_{\hat{\alpha}\hat{ \beta}}$ is not satisfied
along the brane world-volume:
\begin{equation}
\partial_{[\hat{\alpha}} F_{\hat{\beta}\hat{\gamma}]} = H_{\hat{\alpha}\hat{\beta}\hat{\gamma}}\, .
\end{equation}
The anomaly is therefore absent in the presence of $H_{(3)}$ fluxes, only if the flux on 
the world-volume of each separate brane vanishes.  Since a $D9$-brane covers the 
whole ten-dimensional space, its use for stabilizing
the axion-dilaton modulus is therefore ruled out. For $D7$-branes on the other hand, one can 
obtain models free of any FW-anomaly by choosing at least one index
$(\hat{\alpha},\hat{\beta},\hat{\gamma})$ of $H_{\hat{\alpha}\hat{\beta}\hat{\gamma}}$ along the directions transverse to the 7-brane world-volume. 
 
 Furthermore, similar to the toroidal compactification of Section \ref{D9setup}, the number 
 of magnetized $D7$-branes must be doubled. Indeed, the orientifold action 
 ${\cal O}_2$  maps the fluxes $F$ on a stack of $D7$-branes to its inverse 
 ${\cal O}_2: F \rightarrow -F$. To obtain an invariant configuration, a mirror stack 
 must be added with the opposite flux on their world-volume but the same 
 multiplicity and winding numbers.

\subsection{Supersymmetry, tadpoles and charged scalar VEV's}

Let us assume there exist $K$ stacks of $N_a$ magnetized $D7$-branes, $a=1,\dots, K$. For the sake of simplicity, it is assumed that the covering matrices $W_a$ are diagonal. Their entries are thus the wrapping numbers around four of the 1-cycles of $T^6$. 
\be
W_{\; \hat{\alpha}}^{\alpha , \, a} =n^a_\alpha \; \delta^{\alpha}_{ \; \hat{\alpha}}
\ee
 Each of the $D7$-branes covers a 4-cycle of the torus $T^6$. The winding matrix has therefore rank four and two out of the six entries must vanish. 
Finally, the quantized first Chern numbers $m^a_{\hat{\alpha}\hat{\beta}}$ of the magnetic fluxes are given in eq.~(\ref{stab:gen:F}). 

This configuration of magnetized $D7$-branes with 3-form fluxes must satisfy the tadpole conditions. Generally, magnetized $D7$-branes  generate
5-brane and 3-brane tadpoles in addition to the 7-brane charges. Similarly as in 
Section \ref{D9cons},  the presence of mirror branes cancels the 5-brane charges.  
One is then left with 7-brane and 3-brane tadpole contributions. Together with the contribution of the $O3$ planes and of the 3-form fluxes, the tadpole conditions read
\bea
16 & = & \sum_a N_a  {\cal Q}_{3}^a +N_3
\label{t3}\\
0   & = & \sum_a N_a \; n^a_{\alpha}\, n^a_{\beta}\, n^a_{\gamma}\, n^a_{\delta} \quad \quad, \quad \forall \, \alpha,\cdots,\delta = 1 ,\dots, 6
\label{t7}
\eea
where
\be
{\cal Q}_{3}^a =\epsilon^{\hat{\alpha}\hat{\beta}\hat{\gamma}\hat{\delta}}\; m^a_{\hat{\alpha}\hat{\beta}}\; m^a_{\hat{\gamma}\hat{\delta}}\quad \quad ; \quad  N_3 = -{1\over 2}{1\over (2\pi)^4 \alpha'^2}\int_{T^6}H_{(3)} \wedge F_{(3)}
\label{q3}
\ee
These conditions restrict the rank of the gauge symmetry as well as the chiral spectrum. It ensures in particular via the Green-Schwarz mechanism that the spectrum in anomaly-free.

One may ask whether the different stacks of magnetized $D7$-branes preserve the same supersymmetry. The supersymmetry conditions, for a magnetized $D7$-brane on $T^4$, can be
read from general expressions of the central charge or the Born-Infeld
action. The conditions depend on the wrapping matrices $W_a$, the flux ${\cal F}_a$ and the metric moduli of $T^4$:
\be
F^a_{(2,0)} = 0 \quad ; \quad J \wedge {\cal F}_a = 0 \quad ; \quad \left(\prod_\alpha n^a_{\alpha}\right) \Big(J \wedge J - {\cal F}_a \wedge {\cal F}_a\Big) < 0 \quad , \quad \forall a=1,\dots, K\, .
\label{susy7-3a}
\ee
From the 7-brane tadpole condition (\ref{t7}), it is obvious that one needs stacks of branes
with both positive and negative values for the overall winding to cancel 
$D7$ tadpoles. For negative overall winding number $\prod_\alpha n^a_{\alpha}<0$, the positivity condition of eq.~(\ref{susy7-3a}) can be easily satisfied,
consistently with the supersymmetry requirement $J\wedge {\cal F}_a = 0$. 
However, for positive winding $\prod_\alpha n^a_{\alpha}>0$, 
the second set of conditions are impossible to satisfy simultaneously with the inequality  of (\ref{susy7-3a}). It is therefore not possible to obtain supersymmetric  models
with cancelled tadpoles, purely from magnetized $D7$-branes.

To construct consistent supersymmetric models, one may turn on charged
scalar VEV's, as was done for the magnetized $D9$-branes in Section \ref{D9susy}.
The supersymmetry  conditions then read:
\begin{equation}
F^a_{(2,0)} = 0 \ ; \quad \int_{T^4} J\wedge {\cal F}_a = -v_a \int_{T^4}\left( J\wedge J - {\cal F}_a\wedge {\cal F}_a \right) \label{susy7-3} \ ; \quad \left(\prod_\alpha n^a_{\alpha}\right)\Big( {\cal F}_a\wedge {\cal F}_a -J\wedge J \Big)> 0  \ , \ \forall a
	\end{equation}
where the integration space $T^4$ is the appropriate 4-cycle,
$v_a = \sum_k q^a_k |V_{a,\,k}|^2$, with $q^a_i$ being the charge of the 
$k$-th scalar field acquiring a
VEV $V_{a, \, k}$.  Also, in general, $v_a$ can take different values in different 
stacks. The last inequality is a consequence of the positivity of gauge couplings,
as in the case of magnetized $D9$-branes (\ref{gaugecoupling}):
\begin{equation}
{1\over g_a^2}={1\over (4\pi^2\alpha')^2}\left(\prod_\alpha n^a_{\alpha}\right) 
\int_{T^4}\left({\cal F}_a\wedge {\cal F}_a -J\wedge J \right)\, .
\end{equation}

We now introduce two sets of brane stacks: the first set has negative overall winding numbers 
$\prod_\alpha n^b_{\alpha}<0$, $b=1,\dots, 6$ and no VEV for the charged scalars, 
while the second has positive windings and a non-vanishing VEV for some charged 
scalars. We use supersymmetry conditions (\ref{susy7-3}) and 
explicitly construct a model where all the K\"ahler and complex structure 
moduli are stabilized. Specifically, the model we construct has nine 
stacks. The first six stacks contain purely off-diagonal (oblique) fluxes, i.e. 
components mixing the two $T^2$'s of $T^4$. Among these six stacks, the
fluxes along the 
first three are purely symmetric in complex coordinates $z_i = x^i + \tau y^i$, 
$i=1,2,3$ describing the three tori, whereas the remaining three are 
antisymmetric.
The last three stacks have fluxes purely along the diagonal components,
$F_{i\bar{i}}$. Such fluxes are consistent supersymmetric solutions of the tadpole 
equations provided some scalar
VEV's $V_i$ are also simultaneously turned on. We are able to show
that one can consistently stabilize the K\"ahler moduli as a function of 
$V_i$'s, for $V_i << 1$. We now go on to present the model explicitly.

\subsection{A model}\label{modelD7}

Below, we give an explicit model of magnetized $D7$-branes, where all metric moduli are stabilized. 

\begin{table} 
\vskip-0.5cm
\begin{center}

\begin{tabular}{|c|c|c|c|c|}
\hline
Stack No. & Multiplicity  &Flux  & Fixed moduli &  Windings \cr
 &  &    & &  \cr
\hline

Stack-1 & $N_1 = 1$  & $ p_{x^1y^2} = p_{x^2y^1} = 1$ & $\tau_{11} = \tau_{22} \; ; \; \tau_{13} = \tau_{23}=0$ &
$(1, 1, 1, -1,0,0)$ \cr
\hline

Stack-2 & $N_2 = 1$ & $ p_{x^1x^2} = p_{y^1y^2} = 1$ & 
$\tau_{11} \tau_{22} = -1$ & 
$(1, 1, 1, -1,0,0)$ \cr
\hline

Stack-3 &  $N_3=1$ & $ p_{x^2y^3} = p_{x^3y^2} = 1$ & $\tau_{22} = \tau_{33}  \; ; \; \tau_{21} = \tau_{31}=0 $& 
$(0,0, -1, 1,1,1)$ \cr
\hline

Stack-4 & $N_4=1$ & $ p_{x^2x^3} = p_{y^2y^3} = 1$ & $\tau_{22} \tau_{33} = -1$&
$(0,0,-1,1,1,1)$ \cr 
\hline

Stack-5 & $N_5=1$ & $ p_{x^3y^1} = p_{x^1y^3} = 1$ &  $\tau_{11} = \tau_{33} \; ; \; \tau_{12} = \tau_{32}=0  $ &
$(1,1,0,0,1,-1)$ \cr
\hline

Stack-6 & $N_6=1$ &  $ p_{x^3x^1} = p_{y^3y^1} = 1$ & $\tau_{33} \tau_{11} = -1$&
$(1,1,0,0,1,-1)$ \cr
\hline
\end{tabular}
\vskip .4cm
\end{center}
\caption{Brane configuration for the first six stacks of  $D7$-branes with oblique magnetic fluxes.}
\label{table-3}
\end{table}

The first six stacks have multiplicity $N_b=1$, $b=1,\dots, 6$ and  a negative 7-brane charge. 
Their fluxes are all oblique  (given in Table \ref{table-3}). These then fix all metric moduli to be at a point of the moduli space where the six-dimensional torus is a direct product of three squared tori $T^6 = T^2 \times T^2 \times T^2$, or 
\be
\tau_{ij} = i \delta_{ij} \quad  \quad ; \quad \quad J = J_{i\bar{i}} \; dz^i \wedge d\bar{z}^{\bar{i}}\quad \quad , \quad \quad J_{i\bar{j}} = 0
\ee
while the K\"ahler moduli of the three $T^2$ remain unfixed. 
In addition to the above six branes, three more stacks are added
where  charged scalar VEV's are turned on. For these
stacks fluxes and windings are given in Table \ref{table-4}. Let us assume for the sake of simplicity that all charged fields acquire the same VEV $v_a \equiv v$, $a=7,8,9$.
\begin{table}
\vskip .4cm
\begin{center}
\begin{tabular}{|c|c|c|c|}
\hline
Stack No. & Multiplicity & Flux  & Windings \cr
 & & &  \cr
\hline

Strack-7 & $N_7=2$ &   $ p_{x^1y^1} = p_{x^2y^2} = 1$ &   
$(1,1,1,1,0,0)$ \cr
\hline

Stack-8 & $N_8=2$ & $ p_{x^2y^2} = p_{x^3y^3} = 1$ &
$(0,0,1,1,1,1)$ \cr
\hline

Stack-9 & $N_9=2$ & $p_{x^1y^1} = p_{x^3y^3} = 1$ &
$(1,1,0,0,1,1)$ \cr
\hline
\end{tabular}
\vskip .3cm
\end{center}
\caption{Brane configuration for the last three stacks of $D7$-branes with diagonal magnetic fluxes and charged scalar VEV's. }
\label{table-4}
\end{table}
The fluxes are diagonal and fix the remaining three K\"ahler moduli in terms of the 
magnetic fluxes and the VEV of the charged fields. These are constrained by the 
second set of conditions of eq.~(\ref{susy7-3}) to  
\begin{equation}
(J_{1\bar{1}} + J_{2\bar{2}})  = -v(J_{1\bar{1}}J_{2\bar{2}} - 1)
\label{susy7}
\end{equation}
\begin{equation}
(J_{2\bar{2}} + J_{3\bar{3}})  = -v(J_{2\bar{2}}J_{3\bar{3}} - 1)
\label{susy8}
\end{equation}
\begin{equation}
(J_{3\bar{3}} + J_{2\bar{2}})  = -v(J_{3\bar{3}}J_{1\bar{1}} - 1)\, .
\label{susy9}
\end{equation}
These equations have a unique solution:
\begin{equation}
J_{1\bar{1}} = J_{2\bar{2}} = J_{3\bar{3}} = J\, ,
\label{Js}
\end{equation}
with 
\begin{equation}
J = - {1\over v}\left(1 - (1 + v^2)^{\frac{1}{2}} \right)\, .
\label{J-answer}
\end{equation}
For small $v$, the positivity conditions of eq.~(\ref{susy7-3}) are satisfied and one has
\begin{equation}
J \simeq \frac{v}{2}
\label{J-approx}
\end{equation}
for $v$ positive. The sign of $v$ and consequently the positivity of the volume $J$ can be easily verified by the presence of a tachyonic state in the spectrum in the limit where the VEVs vanish.    We  have therefore shown that the above 9 stacks stabilize all 
K\"ahler moduli.

\subsection{Tadpole cancellation}\label{tad73}

The non-zero 7-brane tadpole contributions for various branes may be computed with the flux quanta given in Tables \ref{table-3} and \ref{table-4}. One can easily check that all 7-brane tadpoles vanish, while the 3-brane tadpoles add up to

\begin{equation}
	{\cal Q}_3^{tot} = \sum_{a=1}^9 N_a {\cal Q}_3^a = 12\, ,
\end{equation}
with ${\cal Q}_3^a = 1$ for each brane.

\section{Dilaton Stabilization}

\subsection{3-form fluxes}

A possible stabilization mechanism for the dilaton is by turning on  
R-R and NS-NS 3-form closed string fluxes, that for generic 
Calabi-Yau 
compactifications can fix also the complex structure 
\cite{Frey:2002hf}. 
As we are going to combine the two mechanisms, namely
magnetic and 3-form fluxes, to stabilize the
axion-dilaton modulus for our model in Section 4, we first review 
briefly the main properties of 3-form fluxes. 

Let $H_{(3)}$ and $F_{(3)}$ be the field strengths of the NS-NS 
2-form 
$B_{(2)}$ and of the R-R 2-form $C_{(2)}$, respectively, 
$H_{(3)}=dB_{(2)}$ and
$F_{(3)}=dC_{(2)}$, subject to the usual Dirac quantization 
condition in the compact space. In the basis $(\alpha_{a},\beta_{b})$ 
given in eq.~(\ref{H3basis}) of Appendix \ref{notations}, $H_{(3)}$ and $F_{(3)}$ 
can be written as 
\bea
\frac{1}{(2\pi)^{2}\alpha'}H_{(3)} & = & 
\sum_{a=0}^{h_{2,1}}\left(h^{a}_{1}\alpha_{a} + 
h^{a}_{2}\beta_{a}\right) \nonumber \\
\frac{1}{(2\pi)^{2}\alpha'}F_{(3)} & = &\sum_{a=0}^{h_{2,1}}\left( 
f^{a}_{1}\alpha_{a} + 
f^{a}_{2}\beta_{a}\right),
\label{hf_quanta}
\eea
where  $h_{1}^{a}$, $h_{2}^{a}$, $f_{1}^{a}$ and $f_{2}^{a}$ are 
integers. Using the complex dilaton modulus, one can then form the 
3-form $G_{(3)}$  
\be
 G_{(3)} = F_{(3)}- \phi H_{(3)} \quad , \quad \phi = 
C_{(0)} + i g_{s}^{-1}, \label{gfh} 
\ee
where $g_s$ is the string coupling.  The 3-form background fields 
preserve then a common supersymmetry with the 
$\mathbb{Z}_2$-orientifold projection of $T^6/\mathbb{Z}_2$  if the 
following  
conditions are fulfilled: $G_{(3)}$ has to be a primitive $(2,1)$ 
form \cite{Grana:2001xn}:
\be
 G_{(3)}\wedge J = 0 \quad , \quad  {G_{(3)}} \in H^{2,1}.
\label{primitivity}
\ee
Actually, the second of the conditions above corresponds to finding a 
minimum of the GVW 
superpotential \cite{Gukov:1999ya}
\be
W = \int_{T^{6}} G_{(3)} \wedge \Omega\, ,
\label{supot}
\ee
with $\Omega$ the holomorphic 3-form (\ref{omega}).
$W$ has then to be covariantly constant with respect to all moduli, 
$D_IW = 0$,
or equivalently:
\be
W = 0 \quad , \quad \partial_{\phi}W = 0 \quad , \quad 
\partial_{\tau_{ij}}W = 0,
\label{supotcond}
\ee
where $\phi$ is defined in (\ref{gfh}).  Note that all primitive 
$(2,1)$-forms are 
imaginary self-dual (ISD), $\star_6 G_{2,1} = i G_{2,1}$, where the 
star map $\star_6$ is the usual Hodge map on the torus. 

Let us analyze further the supersymmetry conditions 
(\ref{supotcond}). For given flux quanta (\ref{hf_quanta}), they can 
be understood as conditions on the dilaton and complex structure 
moduli. More precisely, using the symplectic structure 
(\ref{symp_st}), the superpotential (\ref{supot}) reads 
\be
W=\frac{1}{(2\pi)^{2}\alpha'} \int_{T^6} G_{(3)}\wedge \Omega  = 
-(f_{1}^{0}-\phi h_{1}^{0})\det\tau  + 
(f_{2}^{0}-\phi h_{2}^{0}) + (f_{1}^{ij}-\phi h_{1}^{ij})({\rm 
cof}\tau)_{ij} + (f_{2}^{ij}-\phi h_{2}^{ij})\tau_{ij}.
\ee
We can now express the three supersymmetry conditions 
(\ref{supotcond}) explicitly in the form :  
\bea
0 & = & -(f_{1}^{0}-\phi h_{1}^{0})\det\tau  + 
(f_{2}^{0}-\phi h_{2}^{0}) + (f_{1}^{ij}-\phi h_{1}^{ij})({\rm 
cof}\tau)_{ij} + (f_{2}^{ij}-\phi h_{2}^{ij})\tau_{ij} 
\label{3ffcond1}\\
0 & = & h_{1}^{0}\det\tau  - h_{2}^{0} - h_{1}^{ij}({\rm 
cof}\tau)_{ij} - h_{2}^{ij}\tau_{ij} \label{3ffcond2}\\
0 & = & -(f_{1}^{0}-\phi h_{1}^{0}) ({\rm cof}\tau)_{kl} + 
(f_{2}^{kl}-\phi h_{2}^{kl}) + (f_{1}^{ij}-\phi h_{1}^{ij}) 
\epsilon_{ikm}\epsilon_{jln}\tau_{mn}, \label{3ffcond3}
\eea
where ${\rm cof}\tau = (\det\tau)\tau^{-1,T}$. These are eleven 
conditions 
on the complex structure, 
parametrized by the nine elements $\tau_{ij}$ and the 
(complex) dilaton field $\phi$. It is then in principle 
possible to fix all 
complex structure and dilaton moduli in terms  of adequate 
quanta \cite{Frey:2002hf}. Let us now examine the 
primitivity condition $G_{(2,1)}\wedge J = 0$. We could naively think 
that this  can be interpreted, for given fluxes,  as 
conditions on the K\"ahler moduli. However, this condition is 
trivially satisfied in the case of generic Calabi-Yau 
compactifications, because 
there are no harmonic $(3,2)$ forms on these manifolds.  
Therefore, this condition can only become partially non-trivial  on 
 K\"ahler moduli for compactification  manifolds with 
more symmetries, such as the torus. 

There exist however alternative possibilities to fix the metric 
moduli.  As shown in previous sections, the presence of 
internal magnetic fluxes leads to conditions on both the K\"ahler 
class and complex structure moduli. 
For generic Calabi-Yau spaces one can fix only the former\footnote{Note however
that one can also fix complex structure moduli when non-trivial ``fluxes" are
turned on for scalar fields or for gauge field components with no physical zero
modes \cite{AM}.}, while
for toroidal compactifications it is possible to fix all metric 
moduli by a suitable choice of stacks of magnetized $D9$ 
or $D7$-branes. 
On the other hand, the dilaton modulus remains unfixed, but can be stabilized using
3-form closed string fluxes. In fact, for fixed complex structure, 
the conditions 
(\ref{3ffcond1}) and (\ref{3ffcond3}) constrain 
exclusively the dilaton. Moreover, as the K\"ahler form is fixed 
by the presence of magnetic fields, 
the primitivity condition $G_{(2,1)}\wedge J = 0$ restricts 
the possible fluxes $G_{(2,1)}$ we can switch on. Finally, the value 
of the string coupling we can obtain in this way is strongly 
constrained by the tadpole conditions. 
The latter can be read off from the topological coupling of the 
3-form 
fluxes with the R-R 4-form $C_{(4)}$ potential in the 
effective action of the ten-dimensional type IIB supergravity:
\be
S_{CS} = \frac{1}{4i (2\pi)^7 \alpha'^4}\int_{\mathcal{M}_{10}} 
\frac{C_{(4)} \wedge G_{(3)}\wedge \bar{G}_{(3)} }{{\rm Im}\phi} = 
-\mu_3 \, \frac{1}{2} {1\over (2\pi)^4 
{\alpha'}^2}\int_{\mathcal{M}_{10}}C_{(4)}\wedge H_{(3)}\wedge 
F_{(3)},
\label{IIBCS}
\ee 
where we defined the R-R charge $\mu_3$ in terms of $\alpha'$ as 
$\mu_3 = (2\pi)^{-3}{\alpha'}^{-2}$.
The coupling of $C_{(4)}$ to the magnetized $D9$-branes 
gives its effective $R-R$-charge, while the coupling of the $O3$ orientifold 
plane reads
\be
S_{O3} = \mu_3 Q_{O3} \int_{\mathcal{M}_4}C_{(4)}\, ,
\label{EffActO3}
\ee
where $Q_{O3}$  is the R-R-charge of the $O3$ plane.
Therefore, the integrated Bianchi identity for 
the modified R-R 5-form field strength $F_{(5)}$ reads
\be
 -\frac{1}{2}\frac{1}{(2\pi)^{4}\alpha'^{2}}\int_{T^{6}}H_{(3)}\wedge 
F_{(3)} + {\cal Q}^{tot}_{3} + Q_{O3} = 0,
\label{N3}
\ee
where the factor $1\over 2$ comes from the fact that the volume of 
the orientifold 
$T^6/\mathbb{Z}_2$ is half the volume of the torus 
$T^6$.\footnote{Note that it does not come from the factor 
${1\over 2}$ in (\ref{IIBCS}) which is compensated by the magnetic 
coupling 
to $C_{(4)}$; see \cite{Giddings:2001yu} for more details. }  
 
It follows from the ISD condition, that the contribution to (\ref{N3})
coming from the 3-form flux is always positive :
\be
N_3 
=:-\frac{1}{2}\frac{1}{(2\pi)^{4}\alpha'^{2}}\int_{T^{6}}H_{(3)}\wedge 
F_{(3)}= \frac{1}{2g_s} \int_{T^6} H_{(3)}\wedge \star_6 H_{(3)}>0.
\label{N3pos}
\ee
 Finally, the 3-brane
tadpole could also receive contributions from ordinary $D3$-branes. 
All together, the tadpole condition is now modified as 
\be
N_3 + {\cal Q}^{tot}_{3} + N_{D3} + Q_{O3} = 0,
\label{tad3ff}
\ee
where $Q_{O3} = -16$. As the first three terms in the l.h.s. of 
equation
(\ref{tad3ff}) contribute positively, the possible values of  $N_3$ 
as well of ${\cal Q}^{tot}_{3}$ are  bounded. This restricts strongly the 
possible 
values of the string coupling $g_s$. In order to obtain a small value
for the string coupling, $N_3$ should be as large as possible and 
a small contribution from the integral 
$\int_{T^6}H_{(3)}\wedge \star_6 H_{(3)}$. This depends on the quanta 
$h_1^a$ , $h_2^a$ of (\ref{hf_quanta}) and on the Hodge star 
operator. 
The latter only depends on the complex structure 
\cite{Ceresole:1995ca}. It is therefore in principle possible to fix 
the string coupling $g_s$ at small value with the 
help of either internal magnetic fields or 3-form fluxes. This 
is discussed in the next sub-section.

\subsection{Application to the model of Section \ref{D7}}

To stabilize the dilaton, consistently with the geometric moduli stabilization 
achieved by magnetized branes in Section \ref{D7}, 
we now introduce 3-form NS-NS and R-R fluxes to saturate
the total 3-brane tadpole to 16. As mentioned in Section \ref{tad73},
each magnetized $D7$-brane in our model of Section \ref{modelD7} 
contributes one unit in the effective 3-brane charge, 
so that that the nine magnetized 7-brane stacks,  stabilizing all other 
close string moduli, already contribute ${\cal Q}^{tot}_3 = 12$.  
They also fix the complex structure to $\tau_{ij} = i\delta_{ij}$. We 
therefore look for a 3-form flux solution with $N_3 = 4$.

We thus introduce the minimal 
3-form flux which is free of FW-anomalies by taking:
\begin{equation}
h_1^0 \quad , \quad  f_2^0 \quad , \quad h_2^{11} \quad , \quad f_1^{11} 
\end{equation}
for $H_3$ and $F_3$ given in eqs.~(\ref{hf_quanta}).
The supersymmetry conditions eqs.~(\ref{3ffcond1}) - (\ref{3ffcond3}) then
imply:
\begin{equation}
h_1^0 =-h_2^{11} \quad  ; \quad f_2^0 = f_1^{11} \quad \quad ; 
\quad \quad g_s = {h_1^0 \over f_2^0} = -{h_2^{11}\over f_1^{11}}
\end{equation}
and the induced 3-brane charge is
\begin{equation}
N_3 = {1\over 2} \int_{T^6} H_3 \wedge F_3 = 
{1\over 2}(h_1^0 f_2^0 - f_1^{11} h_2^{11}) = h_1^0 f_2^0\, .
\end{equation}
It follows that the minimal value for the string coupling is obtained by 
$h_1^0 =1 \; , \; f_2^0=4$, such that the 3-brane tadpole 
condition (\ref{tad3ff}) is saturated and $g_s = {1\over 4}$.
Also, for the above choice of flux quanta, $G_{(3)}$ as defined
in eq.~(\ref{gfh}) has the form:
\begin{equation}
 G_{(3)} = - 2i dz_1\wedge (dz_2\wedge d\bar{z}_3 - 
				dz_3\wedge d\bar{z}_2),
\label{g3-explicit}
\end{equation}
which is a $(2, 1)$-form as expected and satisfies the primitivity
condition $J\wedge G = 0$ for $J = \sum_i J_{i\bar{i}} dz_i\wedge d\bar{z}_i$.

\section{Conclusions}

We have therefore shown the stabilization of all geometric closed string 
moduli by using gauge field fluxes and charged scalar VEV's. Moreover,
we have shown that the above stabilization method can be consistently 
implemented together with the ones using 3-form fluxes, in order to
stabilize the dilaton-axion modulus, as well. 
A drawback of the construction is that the charged scalar VEV's,
spoil the exact nature of the string construction, although they do not

influence in general the geometric moduli stabilization per se, which
can be done in the absence of such VEV's as it was illustrated in explicit
examples.
Moreover, the string effective action description 
remains valid, as long as the inclusion of charged scalar effects are kept smaller
than the string corrections whose strength is given by the inverse string tension.
For this reason, we have required that the scalar VEV's, $V_i$'s, 
(with $v = \sum_i q_i |V_i|^2$), are small. Also, one has to ensure that 
the string coupling $g_s$ is stabilized at a small value. 
Our solution satisfies this condition as well.

In the case of a generic Calabi-Yau compactification,
the effective potential of the model has two contributions: 
3-form closed string fluxes generate
$F$-term potential for the complex structure moduli and dilaton-axion, 
whereas magnetic open string fluxes along branes (including the charged scalar
VEV's) generate a $D$-term potential for the K\"ahler moduli.
Both these terms are separately stabilized to zero value, giving a 
Minkowski vacuum. 

An interesting exercise would be to find out if the fluxes along 
the branes can give rise to $AdS_4$ minima as well. Since the fluxes
contribute to the space-time energy-momentum tensor, one can hope
to obtain such a vacuum even when the charged scalar VEV's are 
tuned to zero value. $AdS$ branes have in fact been studied in 
different contexts \cite{skenderis-taylor}. 
It remains to be seen though, whether one is able
to obtain such backgrounds in the presence of magnetic fluxes for toroidal 
compactifications.

Another interesting question is to combine this method with $D$-brane model
building and study properties of the corresponding effective field theory.

\section*{Acknowledgements}
We would like to thank M. Bianchi, F. Denef, E. Dudas, F. Marchesano and  E. Trevigne
for useful discussions. This work was supported in part by the European Commission under the
RTN contracts MRTN-CT-2004-503369, MRTN-CT-2004-005104,
the European Union Excellence Grant MEXT-CT-2003-509661,
CNRS PICS no. 2530 and 3059, by the
Agence Nationale pour la Recherche, France and in part by the INTAS contract 03-51-6346.

\appendix

\section{Appendix: Parametrization of $T^6$ }\label{notations}

Consider a six-dimensional torus $T^{6}$ having six coordinates 
with periodicity normalized to unity 
$x^i=x^i+1$, $y_i=y_i+1$~\cite{Moore:1998pn}. Writing the 
coordinates as $x^i$ , $y_i$, $i=1,2,3$, we choose then the 
orientation\footnote{This is  the orientation of \cite{Moore:1998pn}, 
which is different from the one of \cite{Frey:2002hf}. } 
\be
\int_{T^6} dx^1\wedge dy_1\wedge dx^2\wedge dy_2\wedge dx^3\wedge 
dy_3  = 1
\label{orientation}
\ee
and define the basis of the cohomology $H^3(T^6,\mathbb{Z})$
\bea
\alpha_0 & = &  dx^1\wedge dx^2\wedge dx^3 \nonumber \\
\alpha_{ij} & = & \frac{1}{2}\epsilon_{ilm}dx^l\wedge dx^m\wedge dy_j 
\label{H3basis} \\
\beta^{ij} & = & -\frac{1}{2}\epsilon^{jlm}dy_l\wedge dy_m\wedge dx^i 
\nonumber\\
\beta^0 & = & dy_1\wedge dy_2\wedge dy_3, \nonumber
\eea
forming a symplectic structure on $T^6$:
\be
\int_{T^6} \alpha_a \wedge \beta^b = -\delta_a^b\, \, ,
\,\,\,\, \textrm{for}\,\,  a,b =1,\cdots,h_3/2\, , 
\label{symp_st}
\ee
with $h_3 = 20$, the dimension of the cohomology 
$H^3(T^6,\mathbb{Z})$. 

We can also choose complex coordinates 
\be z^i = x^i + \tau^{ij}y_j,
\label{complex_structure}
\ee 
where $\tau^{ij}$ is a complex $3 \times 3$ matrix 
parametrizing the complex structure. In  this basis, the cohomology 
$H^3(T^6,\mathbb{Z})$ decomposes in four different cohomologies 
corresponding to the purely holomorphic parts and those with mixed 
indices:
\be
H^3(T^6) = H^{3,0}(T^6)\oplus 
H^{2,1}(T^6)\oplus H^{1,2}(T^6)\oplus 
H^{0,3}(T^6).
\ee
The purely holomorphic cohomology $H^{3,0}$ is one-dimensional and is 
formed by the holomorphic three-form $\Omega$ for which we choose the 
normalization
\be
\Omega = dz^1\wedge dz^2 \wedge dz^3. \label{omega}
\ee
In terms of the real basis (\ref{H3basis}), this can be written as 
\be
\Omega = \alpha_0 + \tau^{ij}\alpha_{ij} -{{\rm cof} 
\tau}_{ij}\beta^{ij} + \det \tau \beta_0,
\label{omegareal}
\ee
where ${{\rm cof}\tau}_{ij}$ is given by ${{\rm cof}\tau} = 
\left(\det \tau \right)
\tau^{-1,T}$.
We can then define the periods of the holomorphic 3-form to be
\be
\tau^a = \int_{A_a}\Omega\quad , \quad F_b = \int_{B^b} \Omega\, .
\ee
 Note that the period $F_b$ can be written as the derivative of a 
prepotential $F$: $F_b = \partial_{\tau^b} F$.

Similarly, the cohomology $H^2(T^6,\mathbb{Z})$ decomposes also in 
three cohomologies
\be
H^2(T^6) = H^{2,0}(T^6)\oplus 
H^{1,1}(T^6)\oplus 
H^{0,2}(T^6).
\ee
We choose the basis $e^{i\bar{j}}$ of $H^{1,1}$ to be of the form 
\be
e^{i\bar{j}} = i dz^i \wedge dz^{\bar{j}}.
\label{2fbasis}
\ee
The K\"ahler form can therefore by parametrized as 
\be
J= J_{i\bar{j}} e^{i\bar{j}}.
\ee
As the K\"ahler form is a real form, its elements satisfy the reality 
condition $J^{\dagger}_{i\bar{j}} = J_{j\bar{i}}$. Therefore $J$ 
depends only on nine real parameters.

 \bibliographystyle{fullsort}
  \bibliography{bibliography} 

\begin{thebibliography}{99}

\bibitem{Frey:2002hf}
A.~R.~Frey and J.~Polchinski,
Phys.\ Rev.\ D {\bf 65} (2002) 126009
[arXiv:hep-th/0201029];
S.~Kachru, M.~B.~Schulz and S.~Trivedi,
JHEP {\bf 0310} (2003) 007
[arXiv:hep-th/0201028].

\bibitem{Review}
For a recent review see S.~P.~Trivedi, talk in stings 2004,
and references therein.

\bibitem{AM}
I.~Antoniadis and T.~Maillard,
Nucl.\ Phys.\ B {\bf 716}  (2005) 3
[arXiv:hep-th/0412008].

\bibitem{Blumenhagen:2003vr}
R.~Blumenhagen, D.~Lust and T.~R.~Taylor,
Nucl.\ Phys.\ B {\bf 663} (2003) 319
[arXiv:hep-th/0303016].

\bibitem{Cascales:2003zp}
J.~F.~G.~Cascales and A.~M.~Uranga,
JHEP {\bf 0305} (2003) 011
[arXiv:hep-th/0303024].

\bibitem{Bianchi:2005yz}
 M.~Bianchi and E.~Trevigne,
  arXiv:hep-th/0502147 and
arXiv:hep-th/0506080.

\bibitem{Freed:1999vc}
  D.~S.~Freed and E.~Witten,
  arXiv:hep-th/9907189.

\bibitem{AKM}
  I.~Antoniadis, A.~Kumar and T.~Maillard,
  arXiv:hep-th/0505260.
  
\bibitem{Antoniadis:1997mm}
M.~Cvetic, G.~Shiu and A.~M.~Uranga,
Nucl.\ Phys.\ B {\bf 615}  (2001) 3
[arXiv:hep-th/0107166];
I.~Antoniadis, E.~Gava, K.~S.~Narain and T.~R.~Taylor,
Nucl.\ Phys.\ B {\bf 511} (1998) 611
[arXiv:hep-th/9708075].

\bibitem{Bachas:1995ik}
C.~Bachas,
arXiv:hep-th/9503030.
 
\bibitem{ld}
I.~Antoniadis,
Phys.\ Lett.\ B {\bf 246} (1990) 377;
I.~Antoniadis, N.~Arkani-Hamed, S.~Dimopoulos and G.~R.~Dvali,
Phys.\ Lett.\ B {\bf 436} (1998) 257
[arXiv:hep-ph/9804398].

\bibitem{Kachru:2002he}
  S.~Kachru, M.~B.~Schulz and S.~Trivedi,
  JHEP {\bf 0310}, 007 (2003)
  [arXiv:hep-th/0201028].

\bibitem{Maldacena:2001xj}
  J.~M.~Maldacena, G.~W.~Moore and N.~Seiberg,
  JHEP {\bf 0111} (2001) 062
  [arXiv:hep-th/0108100].

\bibitem{Grana:2001xn}
M.~Grana and J.~Polchinski,
Phys.\ Rev.\ D {\bf 65} 126005 (2002)
[arXiv:hep-th/0106014].

\bibitem{Gukov:1999ya}
S.~Gukov, C.~Vafa and E.~Witten,
Nucl.\ Phys.\ B {\bf 584} (2000) 69
[Erratum-ibid.\ B {\bf 608} (2001) 477]
[arXiv:hep-th/9906070];
S.~Gukov,
Nucl.\ Phys.\ B {\bf 574} (2000) 169
[arXiv:hep-th/9911011].

\bibitem{Giddings:2001yu}
S.~B.~Giddings, S.~Kachru and J.~Polchinski,
Phys.\ Rev.\ D {\bf 66} 106006 (2002)
[arXiv:hep-th/0105097].

\bibitem{Ceresole:1995ca}
A.~Ceresole, R.~D'Auria and S.~Ferrara,
Extension,''
Nucl.\ Phys.\ Proc.\ Suppl.\  {\bf 46} 67 (1996)
[arXiv:hep-th/9509160].

\bibitem{skenderis-taylor}
K. Skenderis and M. Taylor,
JHEP {\bf 0206} 025 (2002)
[arXiv:hep-th/0204054].

\bibitem{Moore:1998pn}
G.~W.~Moore,
arXiv:hep-th/9807087.



\bibitem{Angelantonj:2000hi}
M.~Berkooz, M.~R.~Douglas and R.~G.~Leigh,
Nucl.\ Phys.\ B {\bf 480} (1996) 265
[arXiv:hep-th/9606139].
C.~Angelantonj, I.~Antoniadis, E.~Dudas and A.~Sagnotti,
Phys.\ Lett.\ B {\bf 489} (2000) 223
[arXiv:hep-th/0007090].


\bibitem{Marino:1999af}
M.~Marino, R.~Minasian, G.~W.~Moore and A.~Strominger,
JHEP {\bf 0001} (2000) 005
[arXiv:hep-th/9911206].




\end{thebibliography}

\end{document}